\documentclass{mn2e}
\usepackage{graphicx}

\def\br{SN~1999br}

\def\d{SN~1997D}
\def\kms{km s$^{-1}$}

\def\m100{mag/100$^d$}
\def\ni{{$^{56}$Ni}}
\def\c57{{$^{57}$Co}\/}

\def\ti44{{$^{44}$Ti}\/}
\def\r0{{$R_0$}}

\begin{document}

\title[Low Luminosity Type II Supernovae]
{
Low Luminosity Type II Supernovae: \\
Spectroscopic and Photometric Evolution \thanks{Based on observations collected at 
ESO - La Silla (Chile), ESO VLT - Cerro Paranal (Chile), CTIO (Chile), TNG, WHT and JKT (La Palma, Canary 
Islands, Spain) and Asiago (Italy)}
  }
\author[Pastorello et al.]
        {A. Pastorello$^{1,}$$^{2,}$$^{5}$, L. Zampieri$^{2}$, M. Turatto$^{2}$, E. Cappellaro$^{3}$,\and
	 W. P. S. Meikle$^{4}$, S. Benetti$^{2}$, D. Branch$^{5}$, E. Baron$^{5}$,\and F. Patat$^{6}$, 
	M. Armstrong$^{7}$, G. Altavilla$^{2,}$$^{1}$, M. Salvo$^{8}$, M. Riello$^{2,}$$^{1}$\and
	\\
         $^{1}$ Dipartimento di Astronomia, Universit\`a di Padova, Vicolo dell' Osservatorio 2, I-35122 Padova, Italy\\
	 $^{2}$ INAF - Osservatorio Astronomico di Padova, Vicolo dell' Osservatorio 5, I-35122 Padova, Italy\\
	 $^{3}$ INAF - Osservatorio Astronomico di Capodimonte, Via Moiariello 16, I-80131 Napoli, Italy\\
         $^{4}$ Astrophysics Group, Blackett Laboratory, Imperial College, Prince Consort Road, London SW7 2BZ, England, UK\\
         $^{5}$ Department of Physics and Astronomy, University of Oklahoma, 440 W. Brooke St., Norman, OK 73019, USA\\
	 $^{6}$ European Southern Observatory, Karl-Schwarzschild-Strasse 2, D-85748, Garching bei Munchen,
         Germany\\
         $^{7 }$ Rolvenden, Kent, UK\\
	 $^{8}$ Australian National University, Mount Stromlo Observatory, Cotter Road, Weston ACT 2611, Australia\\ }
	 
\date{Accepted .....;
      Received ....;
      in original form ....}


\maketitle

\begin{abstract}

In this paper we present spectroscopic and photometric observations
for four core collapse supernovae (SNe), namely SNe 1994N, 1999br,
1999eu and 2001dc.  Together with SN 1997D, we show that they form a
group of exceptionally low--luminosity events.  These SNe have narrow
spectral lines (indicating low expansion velocities) and low
luminosities at every phase (significantly lower than those of typical
core--collapse supernovae). The very low luminosity
during the $^{56}$Co radioactive decay tail indicates that the mass of
$^{56}$Ni ejected during the explosion is much smaller 
(M$_{Ni} \approx$ 2--8 $\times$ 10$^{-3}$ M$_{\odot}$)
than the average (M$_{Ni} \approx$ 6--10 $\times$ 10$^{-2}$ M$_{\odot}$). 
Two supernovae of this group (SN 1999br and SN 2001dc) were discovered
very close to the explosion epoch, allowing us to determine the
lengths of their plateaux ($\approx$ 100 days) as well as establishing
the explosion epochs of the other, less--completely observed SNe.  It
is likely that this group of SNe represent the
extreme low--luminosity tail of a single continuous distribution of
SN~II--P events. Their kinetic energy is also exceptionally low.
Although an origin from low mass progenitors
has also been proposed for low--luminosity core--collapse SNe,
recent work provides evidence in favour of the high mass
progenitor scenario.
The incidence of these low--luminosity SNe could be as high
as 4--5$\%$ of all type II SNe.

\end{abstract}


\section{Introduction}

It is widely known that the early spectro--photometric evolution of
core--collapse supernovae (CC--SNe) is very heterogeneous (see
e.g. Patat et al., 1994).  The radius of the progenitor is
believed to play a key role in shaping the early light curve.  
H--rich red giant progenitors 
with large initial radii are thought to produce type II plateau supernovae 
(SNe II--P). Their luminosity remains nearly constant for a relatively
long period (plateau phase, lasting $\sim$ 110--130 days), 
during which the hydrogen envelope (in free expansion) starts to recombine, 
releasing its internal energy.
The observed length of the plateau phase depends on the mass of the
hydrogen envelope [Arnett \shortcite{arne80}, Popov \shortcite{popo92}]. \\
On the other hand, the unusual early light curve of SN~1987A with a broad
maximum about 3 months after the explosion is largely attributable to 
its compact progenitor [Woosley et al. \shortcite{woos87}, Arnett
\shortcite{arne87}]. The luminosity at early time is dimmer than expected 
for a ``typical'' SN II--P. Then, with expansion, most of trapped 
energy from radioactive decay of Ni is released and the luminosity rises 
producing a broad maximum in the light curve.
After the plateau phase is finished, all SNe II--P (including SN 1987A)
have a steep drop in luminosity, marking the passage from the
photospheric phase to the nebular one.

The spectroscopic evolution of all SNe II--P is rather homogeneous, 
showing hydrogen
and metal lines with P--Cygni profiles having typical widths ranging
from about 15000 to 3000 \kms~ during the photospheric phase.

A common feature of most light curves of type II SNe is the linear
tail, produced by the radioactive decay of $^{56}$Co into $^{56}$Fe
(0.98 mag/100$^{d}$). While in some cases condensation of dust grains
in the ejecta and ejecta--CSM interaction phenomena may affect the
late--time evolution, there is nevertheless general agreement that the
radioactive tail of a ``typical'' CC--SN is powered by
$\sim$0.06--0.10 $M_\odot$ of $^{56}$Ni [e.g. Turatto et
al. \shortcite{tur90}; Sollerman \shortcite{soll02}].

However, in recent years, systematic observations of non--interacting
SNe in the nebular phase have shown that CC--SNe are also heterogeneous 
at late stages and a more complex picture has emerged as follows:
\begin{itemize}
 \item a few CC--SNe, sometimes called {\sl hypernovae} \cite{iwam98},
       show evidence of exceptionally large Ni masses, 0.3--0.9
       $M_\odot$ [Sollerman et al. \shortcite{soll00}, Patat et
       al. \shortcite{pata01}; see however H\"oflich er al. \shortcite{hoff99}
       who find $\sim$ 0.2$M_\odot$ of ejected $^{56}$Ni assuming asymmetric
       explosions]. In addition, some otherwise
       ``normal'' type II--P SNe have produced unusually bright
       radioactive tails, again implying large amounts of $^{56}$Ni
       [e.g. SN 1992am ($\sim$ 0.3 $M_\odot$), Schmidt et
       al. \shortcite{schm94}].
 \item In most CC--SNe the late--time luminosity is produced by
       0.06--0.10 $M_\odot$ of $^{56}$Ni, e.g. SN 1987A [II pec; Menzies et
       al. (1987); Catchpole et al. (1987); Catchpole et al. (1988);
       Whitelock et al. (1988), Catchpole et al. (1989); Whitelock et
       al. (1989)], SN 1988A [II--P; Ruiz--Lapuente et al. (1990);
       Benetti et al. (1991); Turatto et al. (1993)], SN 1993J [IIb;
       see e.g. Barbon et al. (1995)], and SN 1994I [Ic; Young, Baron
       \& Branch (1995); Richmond et al. (1996) and references
       therein].
 \item A few events exhibit a somewhat lower late--time luminosity,
       such as SN 1991G \cite{blan95}, SN 1992ba \cite{ham03a} and SN 1999gi 
       [Leonard et al. \shortcite{leo02b}; Hamuy \shortcite{ham03a}]. 
       This may be attributable to the ejection of reduced amounts of 
       $^{56}$Ni (0.015--0.04 $M_\odot$). Conversely, it could be that the
       luminosity of these SNe is greater, but the distances
       and$/$or the effects of interstellar absorption have been
       underestimated. This might be the case of SN~1999em [Baron et
       al. (2000); Leonard et al. (2002a); Pooley et al.  (2002); Hamuy
       et al. (2001); Elmhamdi et al. (2003)] for which
       recent Cepheid distance measurements \cite{leon03} suggest
       that the host galaxy distance was previously underestimated 
       and, consequently, the SN luminosity probably higher.   
       The type~II--P SN~1994W \cite{soll98}, while
       having an exceptionally high luminosity during the photospheric
       era, was also particularly faint at late times with an ejected
       $^{56}$Ni mass below 0.015 M$_{\odot}$, 
       depending inversely on the assumed contribution to the luminosity 
       of an ejecta--CSM interaction.
 \item Finally, there is a small group of events having very low
       late--time luminosities. The prototype is SN 1997D (Turatto et
       al., 1998).  As demonstrated in this paper, the SNe of this group are also
       exceptionally faint at early--times.  The other SNe in this
       category which we study here are SN~1994N (Turatto, 1994), SN
       1999br (King, 1999), SN 1999eu (Nakano $\&$ Aoki, 1999) and
       SN~2001dc (Armstrong, 2001). As we shall see, the late--time light curves
       of this group require very small amounts of $^{56}$Ni, at least
       one order of magnitude smaller than in SN~1987A.
\end{itemize}
In a companion paper \cite{zamp03} we performed an analysis of the data
of \d\/ and \br, deriving  information about the nature of the progenitor
stars and the explosion energies.
The early observations of \br\/ allowed us to constrain the models discussed
in previous papers [Turatto et al. \shortcite{tura98};
Chugai $\&$ Utrobin \shortcite{chug00}].
We found that these explosions are under--energetic with respect to a
typical type II SN and that the inferred mass of the ejecta is large (M$_{ej} \geq$ 14--20 M$_
{\odot}$).

In this paper we present the spectroscopic and photometric
observations of the very low luminosity SNe 1999br, 1999eu, 1994N and
2001dc. Together with the \d\/ data [Turatto et
al. \shortcite{tura98}; Benetti et al. \shortcite{bene01}] these make
up almost all that is available for this group of SNe.  The plan of
the paper is as follows: in Sect.~2 we describe the SNe and their
parent galaxies. In particular, we estimate distances, crucial for the
derivation of the luminosity and, in turn, of the $^{56}$Ni mass.  The
observations are summarised in Sect.~3. In Sect.~4 we present
photometric data and in Sect.~5 we analyze light and colour curves,
focusing on the common properties of this group of SNe and making
comparisons with the prototype SN 1997D. In Sect.~6 spectroscopic
observations are presented. In Sect.~7 we discuss the data with
particular focus on the progenitors nature. Sect.~8 is devoted to an
estimate of the frequency of these low--luminosity events.  A short
summary follows in Sect.~9.

Throughout this paper we adopt H$_{0}$ = 65 km s$^{-1}$ Mpc$^{-1}$.
In all the images, North is up, East is to the left and the numbers
label stars of the local calibration sequence.

\section{The SNe and their host Galaxies.}
In Tab. \ref{datagal} we summarize the main observational data for
our sample of SNe and their host galaxies.  For completeness, we
show also the parameters for SN~1997D in this table.

\begin{table*}
\caption{Main observational data for the low--luminosity SNe II and their host galaxies.} \label{datagal}
\scriptsize
\begin{tabular}{|c|c|c|c|c|c|c|c|c|c|c|} \hline
SNe Data & 1997D & & 1999br & & 1999eu & & 1994N & & 2001dc \\ \hline
$\alpha$ (J2000.0) & 04h11m01\fs00 & $\bullet$ & 13h00m41\fs80 & $\triangle$ & 02h46m20\fs79 & $\diamond$ & 10h29m46\fs8 & $\nabla$ & 14h51m16\fs15 & $\ast$\\
$\delta$ (J2000.0) & $-56\degr29\arcmin56\farcs0$ & $\bullet$ & $+02\degr29\arcmin45\farcs8$ & $\triangle$ & $-30\degr19\arcmin06\farcs1$ & $\diamond$ & $+13\degr01\arcmin14^{\prime\prime}$ & $\nabla$ & $+58\degr59\arcmin02.8$ &$\ast$\\
Offset SN-Gal.Nucleus &11$^{\prime\prime}$E, 43$^{\prime\prime}$S & $\bullet$  & 40$^{\prime\prime}$E, 19$^{\prime\prime}$S &$\triangle$ & 23$^{\prime\prime}$E, 157$^{\prime\prime}$S & $\diamond$ & 1$\farcs8$E, 9$\farcs8$N & $\nabla$ &26$^{\prime\prime}$W, 28$^{\prime\prime}$N & $\ast$ \\
Discovery Date (UT) & 1997 Jan 14.15 & $\bullet$ & 1999 Apr 12.4 &$\triangle$ & 1999 Nov 5 & $\diamond$ & 1994 May 10.0 & $\nabla$ & 2001 May 30.96 & $\ast$\\
Discovery Julian Date & 2450462.65 & $\bullet$ & 2451280.9 & $\triangle$ & 2451487.5 & $\diamond$ & 2449482.5 &$\nabla$  & 2452060.46 & $\ast$ \\
Explosion Epoch (JD) & 2450361 & $\times$ & 2451278 & $\times$ & 2451394 & $\times$ & 2449451 & $\times$ & 2452047 & $\times$\\
Discovery Magnitude & 16.3 (Jan 15.05) &$\bullet$ & m=17.5 & $\triangle$ & m=17.3 &$\diamond$  & R=17.5 & $\nabla$ & m=18.5 & $\ast$\\
V(max) & $\leq$19.8 & $\times$ & $\leq$16.8 & $\diamondsuit$ & 17.5 & $\times$ & $\leq$17.3 & $\times$ & $\leq$18.5 & $\times$ \\
Total Extinction $A_{B,tot}$ & 0.089 & $\times$ & 0.102 & $\times$ & 0.113 & $\times$ & 0.169 & $\times$ & 1.7 &  $\times$  \\ \hline \hline
Host Galaxies Data & NGC 1536 & & NGC 4900 & & NGC 1097 & & UGC 5695 & & NGC 5777 & \\ \hline
$\alpha$ (J2000.0) & 04h10m59\fs86 & $\dag$ & 13h00m39\fs13 & $\dag$ & 02h46m19\fs06 & $\dag$ & 10h29m46\fs8 & $\dag$ & 14h51m18\fs55 & $\dag$ \\
$\delta$ (J2000.0)  & $-56\degr28\arcmin49\farcs6$ & $\dag$ & $+02\degr30\arcmin05\farcs3$ & $\dag$ & $-30\degr16\arcmin29\farcs7$ & $\dag$ & $+13\degr01\arcmin05\farcs5$ & $\dag$ & $+58\degr58\arcmin41\farcs4$ & $\dag$ \\
Morph. Type & SB(s)c pec & $\dag$ & SB(rs)c & $\dag$ & SBbSy1 & $\dag$ & S? & $\dag$ & Sbc & $\dag$\\
B Magnitude & 13.15 & $\dag$ & 11.90 & $\dag$ & 10.23 & $\dag$ & 14.66 & $\dag$ & 14.11 & $\dag$ \\
Galactic Extinction $A_{B}$ & 0.092 & $\otimes$ & 0.102 & $\otimes$ & 0.113 & $\otimes$ & 0.169 & $\otimes$& 0.046 & $\otimes$ \\
Diameters & 2'.0 x 1'.4 & $\dag$  & 2'.2 x 2'.1 & $\dag$ & 9'.3 x 6'.3 & $\dag$ & 1'.3 x 0'.5 & $\dag$ & 3'.38 x  0'.45 &$\dag$ \\
$v_{hel}$ (km s$^{-1}$) & 1461 & $\star$ & 968 & $\star$ & 1273 & $\star$ & 2940 & $\star$ & 2140 & $\star$ \\
$\mu$ ($H_{0}$=65 km s$^{-1}$ Mpc$^{-1}$)& 31.29 & $\times$ & 31.19 & $\circ$ & 31.08 & $\times$ & 33.34 & $\times$ & 32.85& $\times$\\ \hline\hline
\end{tabular}

$\star$ LEDA \footnotemark[1];
$\dag$ NED \footnotemark[2];
$\triangle$ \protect\cite{king99};
$\otimes$ \protect\cite{schl98};
$\diamond$ \protect\cite{naka99};
$\nabla$ \protect\cite{tura94};\\
$\ast$ \protect\cite{hurs01};
$\circ$ \protect\cite{giur00};
$\bullet$ \protect\cite{deme97};
$\diamondsuit$ \protect\cite{bene01};
$\times$ this paper. 
\end{table*}

\begin{figure}
\includegraphics[width=8.05cm]{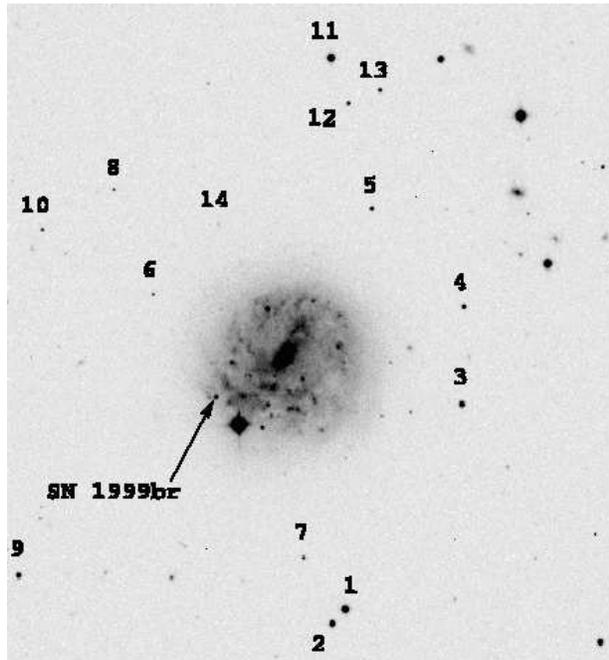}
\caption{V band image of SN 1999br and the host galaxy NGC 4900 (image
obtained on 1999 July 6 with ESO--Danish 1.54m telescope +
DFOSC.)}\label{fig_99br}
\end{figure}

\begin{enumerate}

\item \br\/ was discovered by King \shortcite{king99} in the course of
the Lick Observatory Supernova Search on 1999 April 12.4 UT, and
confirmed the following day, when its magnitude was about 17.5.  It
was located at $\alpha$ = 13h00m41$\fs8$, $\delta$ =
$+02\degr29\arcmin45\farcs8$ (equinox 2000.0), about
40$^{\prime\prime}$ East and 19$^{\prime\prime}$ South from the
nucleus of NGC 4900 and near a bright foreground star (12$\farcs5$
West and 15$\farcs6$ South) of eleventh magnitude (see
Fig. \ref{fig_99br}).  \br\/ is the first 1997D--like event discovered a
few days after the explosion.  There was no evidence of the SN on
frames taken on 1999 March 27.4 UT (limiting magnitude 18.5) and on
1999 April 4.4 (limiting magnitude 17; Li, IAUC 7143).  The SN was
classified as a peculiar, faint type II event [Garnavich et
al. \shortcite{gar99a}, Filippenko et al. \shortcite{fil99b}].  Patat
et al. \shortcite{pata99} pointed out the similarity with SN~1994N at
a comparable phase and suggested that SN~1999br had produced a very
low amount of $^{56}$Ni.\\
The host galaxy (Fig. \ref{fig_99br}) of SN~1999br, NGC~4900, is a
well studied SB(rs)c galaxy lying in the direction of the Virgo
Cluster. Different estimates of the distance have been published for
this galaxy [Bottinelli et al. \shortcite{bott85}, Kraan--Korteveg
\shortcite{kraan86}, Fouqu\`e et al. \shortcite{fouq00}, Ekholm et
al. \shortcite{ekho00}, Freedman et al. \shortcite{free01}].
The recession velocity of NGC 4900 corrected for the Local Group
infall onto the Virgo Cluster (from the LEDA cataloue) is $v_{vir}=1013$ \kms. 
This is close to
the average value for the group, dominated by NGC~4517, to which NGC~4900
belongs [$<v_{vir}>$ = 1125 \kms, Giuricin et al. \shortcite{giur00}],
resulting in a distance modulus of $\mu=31.19$. This is the final value adopted 
for SN~1999br (see Tab. \ref{datagal}).\\ 
For the galactic extinction we adopt A$_B$ = 0.102 \cite{schl98}; no
sign of internal extinction (e.g. lack of narrow interstellar Na ID lines) 
is present in the spectra of SN 1999br,
which is not unexpected given the peripheral location of the SN in
NGC~4900.\\
\footnotetext[1]{http://leda.univ-lyon1.fr/search.html}
\footnotetext[2]{http://nedwww.ipac.caltech.edu/index.html}

\item SN 1999eu was discovered by Nakano and Aoki \shortcite{naka99}
with a 0.40m reflector on 1999 November 5 and confirmed the following
day. The SN was located at $\alpha$ = 02h46m20\fs79, $\delta$ =
$-30$\degr19\arcmin06\farcs1 (equinox 2000.0), 23$^{\prime\prime}$
East, 157$^{\prime\prime}$ South from the nucleus of NGC~1097, lying
on an arm with a relatively flat background structure (see
Fig. \ref{fig_99eu}).  Garnavich et al. (1999b) classified SN~1999eu as
a peculiar SN~II, with a spectrum characterized by several narrow
P--Cygni lines and a typical velocity of $\sim$ 1500 km s$^{-1}$ (from
the minimum of Ba II 6142 A).  Garnavich et al. concluded that
SN~1999eu was an under--luminous type~II event powered by the ejection
of an extremely small amount of $^{56}$Ni.\\ 
The host galaxy, NGC~1097, is a peculiar barred spiral listed in Arp's
Catalogue (1966). Sersic (1973) noted that the galaxy nucleus was
morphologically peculiar with a central condensation, surrounded by an
annulus of hot spots, almost uniformly and symmetrically distributed
(Fig. \ref{annulus}). Wolstencroft et al. (1984) found that this structure was
composed of H~II regions emitting at radio wavelengths.  The
observations suggest that a burst of star formation is taking place in
the ring.  This is supported also by the discovery of the
core--collapse SN~1992bd \cite{schm92} in this region (Fig.  \ref
{annulus}).  The nucleus of the galaxy is a compact radio source and
shows the emission line spectrum of a Seyfert 1 (Storchi--Bergmann et
al., 1997).  We also note that the type~II SN~2003B was discovered in
a peripheral region of the galaxy, close to the nearby small
elliptical companion NGC 1097A \cite{evan03}.  \\
Contrary to what is sometimes claimed, NGC~1097 membership of the
Fornax cluster is uncertain (Giovanelli et al., 1997).  The galaxy
velocity is close to the average value of the group NOG 179, to which
NGC 1097 belongs \cite{giur00}.  The recession velocity corrected for Virgo 
infall was derived from the LEDA catalogue, which yields $v_{vir}=1069$ \kms. 
The resulting distance modulus is $\mu=31.08$ (Tab. \ref{datagal}). \\ 
For the galactic extinction, we adopt A$_B$ = 0.113 \cite{schl98}. There
is no spectroscopic evidence for strong extinction in the host galaxy.

\item SN~1994N was serendipitously discovered by Turatto (1994) with
the ESO 3.6m telescope during an observation of the type~IIn SN~1993N,
which exploded the year before in the same galaxy \cite{mue93a}.
The new SN was located at $\alpha$ = 10h29m46\fs8, $\delta$ =
$+13\degr01\arcmin14^{\prime\prime}$ (equinox 2000.0), 1$\farcs8$ East and
9$\farcs8$ North from the nucleus of the spiral galaxy UGC 5695, not
far (8$\farcs5$ East) from SN~1993N (Fig. \ref{sn94N_fig}).\\ 
Little information on UGC 5695 is available.
It is a member of a small group of galaxies, LGG 207, which has a mean
heliocentric velocity of 2856 \kms~(Garcia, 1993), slightly larger
than that of the galaxy itself.  Taking LEDA's heliocentric velocity
of UGC 5695 corrected for Virgo infall
and for the peculiar motion inside the group (+48 \kms), we derive a
distance modulus $\mu$ = 33.34. \\ 
Because evidence of
strong internal extinction is missing from our spectra, we have applied
only the galactic extinction correction A$_B$=0.169 \cite{schl98}.

\item M. Armstrong \shortcite{hurs01} discovered SN 2001dc at $\alpha$
= 14h51m16\fs15 and $\delta$ = +58$\degr$59$\arcmin$02$\farcs$8
(equinox 2000.0), 26$^{\prime\prime}$ West and 28$^{\prime\prime}$
North of the center of NGC 5777 (see Fig. \ref{sn01dc_field}).  A
prediscovery upper limit indicates that SN~2001dc was discovered only
a few days after explosion. From early photometric monitoring, Meikle
and Fassia (2001, IAU Circ. 7662) noted that this event could be
classified as a SN II--P and concluded that the event was unusually
under--energetic, with an absolute magnitude very low compared with
normal SNe II--P.\\ The host galaxy NGC 5777 is an edge-on Sbc galaxy,
crossed by a spectacular equatorial dust lane \cite{bott90}. From LEDA
we derive a heliocentric velocity of 2140 \kms ~and, after correcting
for Virgo infall, v$_{vir} \approx$ 2419 \kms, $\mu$ = 32.85 (see
Tab. \ref{datagal}).  \\
The galactic extinction is $A_B$ = 0.046 \cite{schl98}. 
The position of the SN, not far from the nucleus and
projected onto a background rich in gas and dust, leads us to suspect
the presence of some internal extinction. We shall discuss this point
in Sect. 5.
\end{enumerate}

\begin{table*}
\caption{Journal of spectroscopic observations (JD +2400000).} \label{obs_sp}
\scriptsize
\begin{tabular}{|c|c|c|c|c|c|} \hline
Date & JD & Phase & Instrument & Range (A) & Res. (A) \\ \hline\hline
\multicolumn{5}{|c|}{SN 1994N} \\ \hline
10/05/94 & 49482.54 & 31.5 & 3.6m+EFOSC1 & 3700--9850 & 13,17  \\  
14/05/94 & 49486.56 & 35.6 & 2.2m+EFOSC2 & 4500--7150 & 11 \\
05/06/94 & 49508.53 & 57.5 & 3.6m+EFOSC1 & 3700--6900 & 18 \\  
30/01/95 & 49747.80 & 296.8 & NTT+EMMI & 3850--8950 & 9 \\ \hline  
\multicolumn{5}{|c|}{SN 1999br} \\ \hline
23/04/99 \footnotemark[3] & 51291.73 & 13.7 & LCO100 & 3700--9300 & 5 \\
26/04/99 \footnotemark[3] & 51294.64 & 16.6 & 1.54m+DFOSC & 3500--9800 & 14,19 \\ 
29/04/99 \footnotemark[3] & 51297.68 & 19.7 & NTT+EMMI & 3350--10250 & 7,6 \\  
03/05/99 \footnotemark[3] & 51301.64 & 23.6 & NTT+EMMI & 3300--10100 & 7,6 \\ 
11/05/99 \footnotemark[3] & 51309.69 & 31.7 & 1.54m+DFOSC & 3700--10100 & 14,19 \\ 
19/05/99 \footnotemark[3] & 51317.68 & 39.7 & NTT+EMMI & 3400--10100 & 7,6 \\ 
21/05/99 & 51320.02 & 42.0 & 3.6m+EFOSC2 & 3350--10250 & 14,17 \\ 
20/07/99 & 51380.45 & 102.5 & 1.54m+DFOSC & 3400--9050 & 12 \\ \hline
\multicolumn{5}{|c|}{SN 1999eu} \\ \hline
10/11/99 & 51492.64 & 98.6 & 1.54m+DFOSC & 3450--9050 & 11  \\
14/11/99 & 51496.81 & 102.8 & 3.6m+EFOSC2 &3400--7500 & 17  \\
05/12/99 & 51517.74 & 123.7 & 3.6m+EFOSC2 & 3350--10250 & 14,17  \\ 
18/12/99 & 51530.71 & 136.7 & 3.6m+EFOSC2 & 3400--7450 & 17  \\ \hline
\multicolumn{5}{|c|}{SN 2001dc} \\ \hline
10/07/01 & 52101.48 & 54.5 & INT+IDS & 4850-9600 & 17  \\
16/08/01 & 52138.44 & 91.4 & TNG+DOLORES & 3300-8000 & 15  \\ 
24/08/01 & 52146.43 & 99.4 & WHT+ISIS & 3750-9950 & 14  \\ \hline\hline
\end{tabular}
\end{table*}

\section{Summary of the Observations}

Our observations of SNe 1999br, 1999eu, 1994N and 2001dc
were obtained with the ESO, CTIO, ING, TNG and Asiago telescopes.
Images and spectra were reduced using standard IRAF or FIGARO
procedures.  The photometric data are presented in Sect. 4.  The
magnitudes of the SNe were obtained using either PSF fitting and
template subtraction techniques, depending on the background
complexity and the availability of suitable template images.\\ The
journal of spectroscopic observations is reported in
Tab. \ref{obs_sp}.  The spectra of SNe 1999eu, 1994N and 2001dc are
presented here for the first time.  All the spectra were flux
calibrated using standard stars [selected from Hamuy et al. (1992),
Hamuy et al. (1994), Stone (1977), Stone $\&$ Baldwin (1983), Baldwin
$\&$ Stone (1984)] observed during the same nights.

The photometric and spectroscopic observations of SN~1999br provide
good coverage of the plateau phase, but only sparse data are available
in the nebular phase. In particular no post--plateau spectra are
available.  The discovery epoch of SN~1999eu was probably long after
the explosion, at the end of the plateau, although it was well
observed both photometrically and spectroscopically at late epochs.
The data for SN~1994N are sparse, although they do span a period of
about 22 months.  The photometric coverage of SN 2001dc began about
one month after discovery. Because of its faintness, it was not
possible to observe it long after the end of the plateau.  Three
spectra were also obtained, all during the plateau phase.
Clearly, the temporal coverage of individual low--luminosity SNe 
is incomplete and erratic. However, we shall argue that these events
form a fairly homogeneous group and so, taken together, they provide
a reasonably complete picture of the photometric and spectroscopic
evolution of this type of supernova.

\begin{figure}
\includegraphics[width=9.5cm]{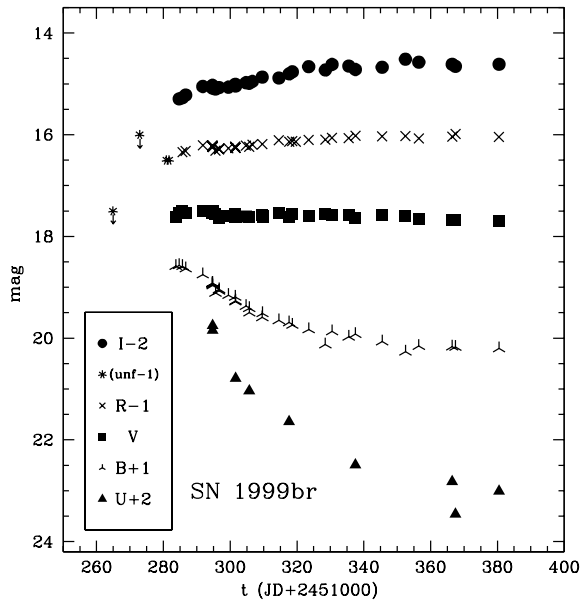}
\caption{UBVRI light curves of SN 1999br showing its extended
plateau. Two prediscovery limits (IAU Circ. 7143), unfiltered
measurements, are also shown placed to correspond with the R magnitude
scale (IAU Circ. 7141). Photometry from Hamuy et al. (in prep.) is also
shown.}
\label{lc_99br}
\end{figure}
\footnotetext[3]{Hamuy $\&$ Phillips, private communication}

\section{Light Curves}

\subsection{SN~1999br}
 
\begin{table*}
\caption{Photometry of SN 1999br (JD +2400000).} \label{sn99br_ph}
\scriptsize
\begin{tabular}{|c|c|c|c|c|c|c|c|} \hline
Date & JD & U & B & V & R & I & Instrument \\ \hline \hline  
 21/05/99 & 51319.65 & -- &  -- & -- &  17.14 (.04) & -- & 1 \\ 
 07/06/99 & 51337.50 & 20.49 (0.29) & 18.92 (0.01)  & 17.63 (0.01)  & 17.02 (0.01)  & 16.72 (0.01) &  2 \\ 
 06/07/99 & 51366.47 & 20.82 (0.37) & 19.17 (0.09) & 17.68 (0.01)  & 17.04 (0.01)  & 16.62 (0.01)& 3 \\ 
 07/07/99 & 51367.45 & 21.46 (0.55) & 19.18 (0.03) & 17.69 (0.04) & 16.98 (0.02)& 16.66 (0.01) &  2 \\ 
 20/07/99 & 51380.48 & 21.01 (0.31) & 19.20 (0.03) & 17.69 (0.01)  & 17.04 (0.02)& 16.62 (0.01) &  3 \\ 
 02/04/00 & 51636.59 & -- &  -- & 22.60 (0.15) & -- & 21.18 (0.24) & 4 \footnotemark[3]\\
 09/04/00 & 51643.69 & -- &  -- & 22.68 (0.08) & -- & -- & 4 \footnotemark[3]\\
 02/05/00 & 51666.67 & -- &  -- & -- & 21.98 (0.05) & -- & 4 \footnotemark[3]\\
 01/02/01 & 51941.87 & -- &  -- & $\ge$24.4  & $\ge$24.6  & -- &  1 \\ 
 26/07/01 & 52117.50 & -- &  -- & $\ge$25.2  & $\ge$24.8  & -- & 5 \\ \hline
\end{tabular}

1 = ESO 3.6m + EFOSC2; 2 = TNG + OIG\\ 
3 = Danish 1.54m + DFOSC; 4 = NTT + EMMI; 5 = ESO VLT + FORS1\\
\end{table*}

The Tololo Group 
performed an intensive follow--up of SN~1999br during the plateau
phase with the CTIO and ESO telescopes (Hamuy, 2003; Hamuy et al.,
in prep.). This SN was also observed by us using the ESO telescopes and
TNG.  Our optical photometry in the UBVRI bands is shown in
Tab. \ref{sn99br_ph}.  The SN magnitudes were calibrated by mean of a
local sequence (Hamuy et al., in prep.) after comparison with Landolt standard
stars \cite{Land92}.
The photometric measurements of SN 1999br were performed using the
template subtraction technique. The template images were obtained on
2000 April 2 (NTT + EMMI) and 2001 February 1 (ESO 3.6m + EFOSC2).
The errors were computed by placing some
artificial stars (having the same magnitude as the SN) at positions
close to the SN, and hence estimated the deviations in the measured
magnitudes.\\
 
The first season UBVRI light curves are shown in Fig. \ref{lc_99br},
which also includes the data from Hamuy et al. (in prep.).  A
prediscovery limit from IAU Circ. 7143 allows us to fix the explosion
epoch between JD = 2451264.9 and 2451280.9. Another limit obtained on
JD = 2451272.9 is less stringent. Hereafter we adopt as the explosion
epoch JD = 2451278 $\pm$3, which is compatible with the assumptions of
Zampieri et al. \shortcite{zamp03}.  While the U and B band light
curves decline monotonically after discovery (the slope of the B band
light curve is 3.66 mag/100$^{d}$ during the first 40 days and 0.61
mag/100$^{d}$ later), the V band light curve shows a plateau of
duration at least 100~days.  Between about days 30 and 100, the slope
is $\gamma_V$ $\approx$ 0.20 mag/100$^{d}$. The R and I band
magnitudes increase with time up to the last data point at
$\sim$100~days.  During the era 30--100 days, the slopes are
respectively $\gamma_R$ $\approx -$0.18 and $\gamma_I$ $\approx -$0.25
mag/100$^{d}$. This evolution is typical of a SN~II during the plateau
phase \cite{popo92}. Unfortunately, no measurements are available
between 1999 July 20, and 2001 February 1, and so the precise length
of the plateau phase is undetermined.  Only a few measurements or
upper limits (see e.g. Fig. \ref{abs_V}) are available during the
radioactive decay epoch, but they are of special interest due to their
role in the determination of the $^{56}$Ni mass.

\begin{figure}{
\includegraphics[width=8.05cm]{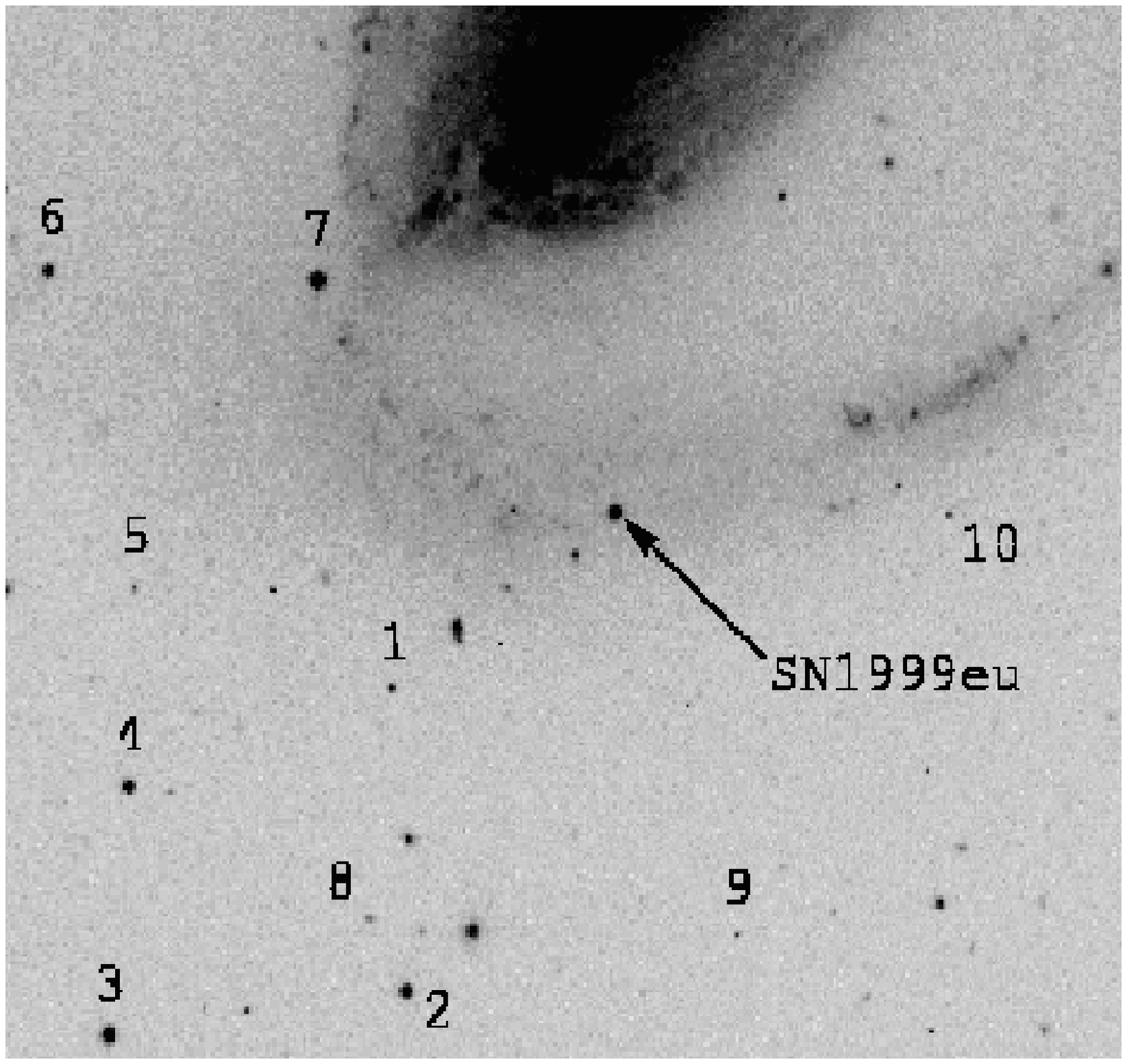}
\caption{SN 1999eu in NGC~1097 (R band image with ESO--Danish 1.54m telescope
         + DFOSC)} \label{fig_99eu}
\hspace{2.5cm}	 
\includegraphics[width=5.7cm]{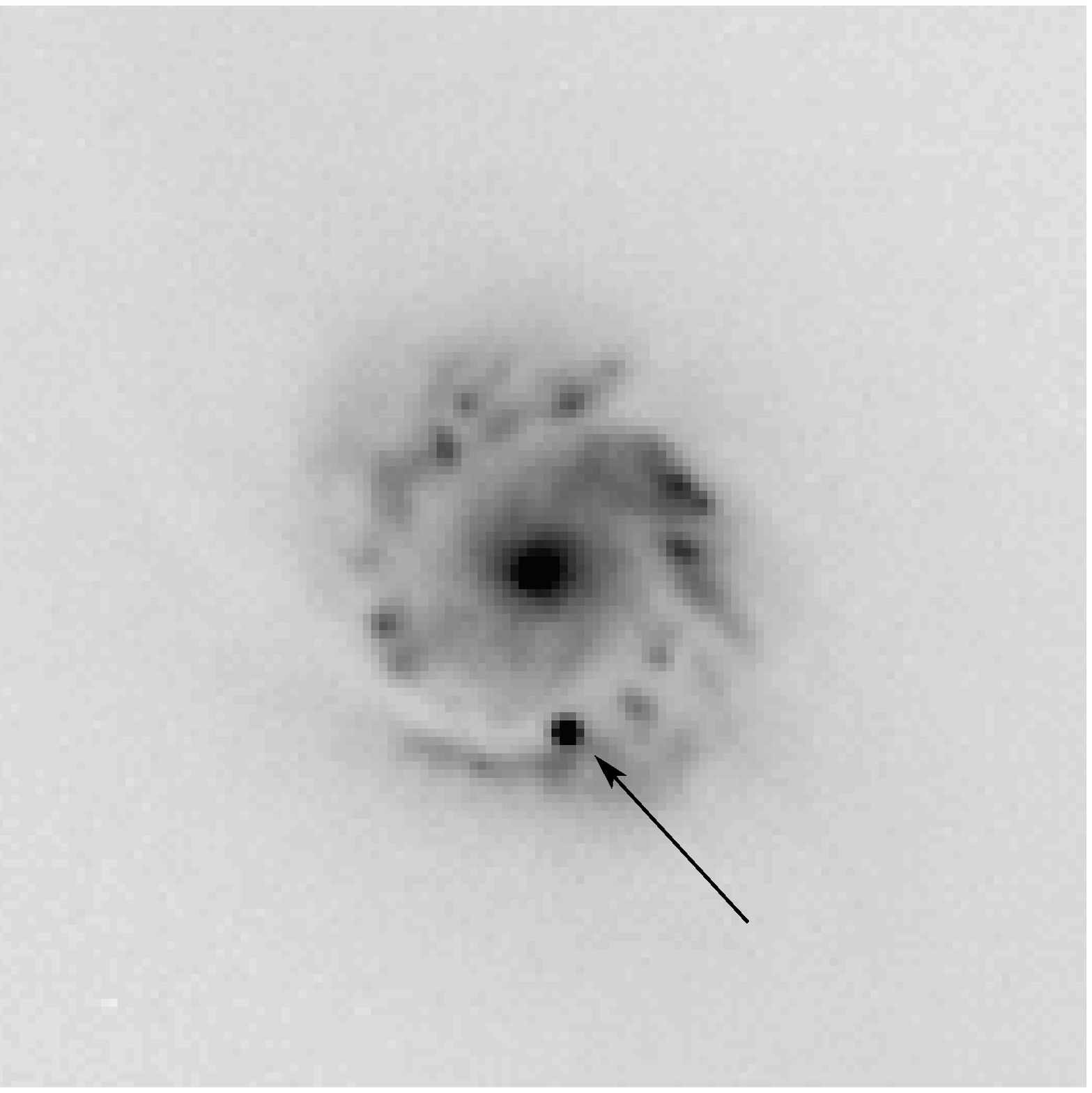}\\
\caption{The annulus of H II regions and hot spots surrounding the
nucleus of NGC 1097, possibly a region of intense star formation. The
frame was taken on 1992 October 16 with NTT + EMMI (filter R) 
and includes the type~II SN~1992bd,
indicated by an arrow. NGC~1097 also hosted SN~1999eu, studied in this
paper, as well as the type~II SN~2003B.}
\label{annulus} 
}
\end{figure}

\begin{figure}
\includegraphics[width=9.0cm]{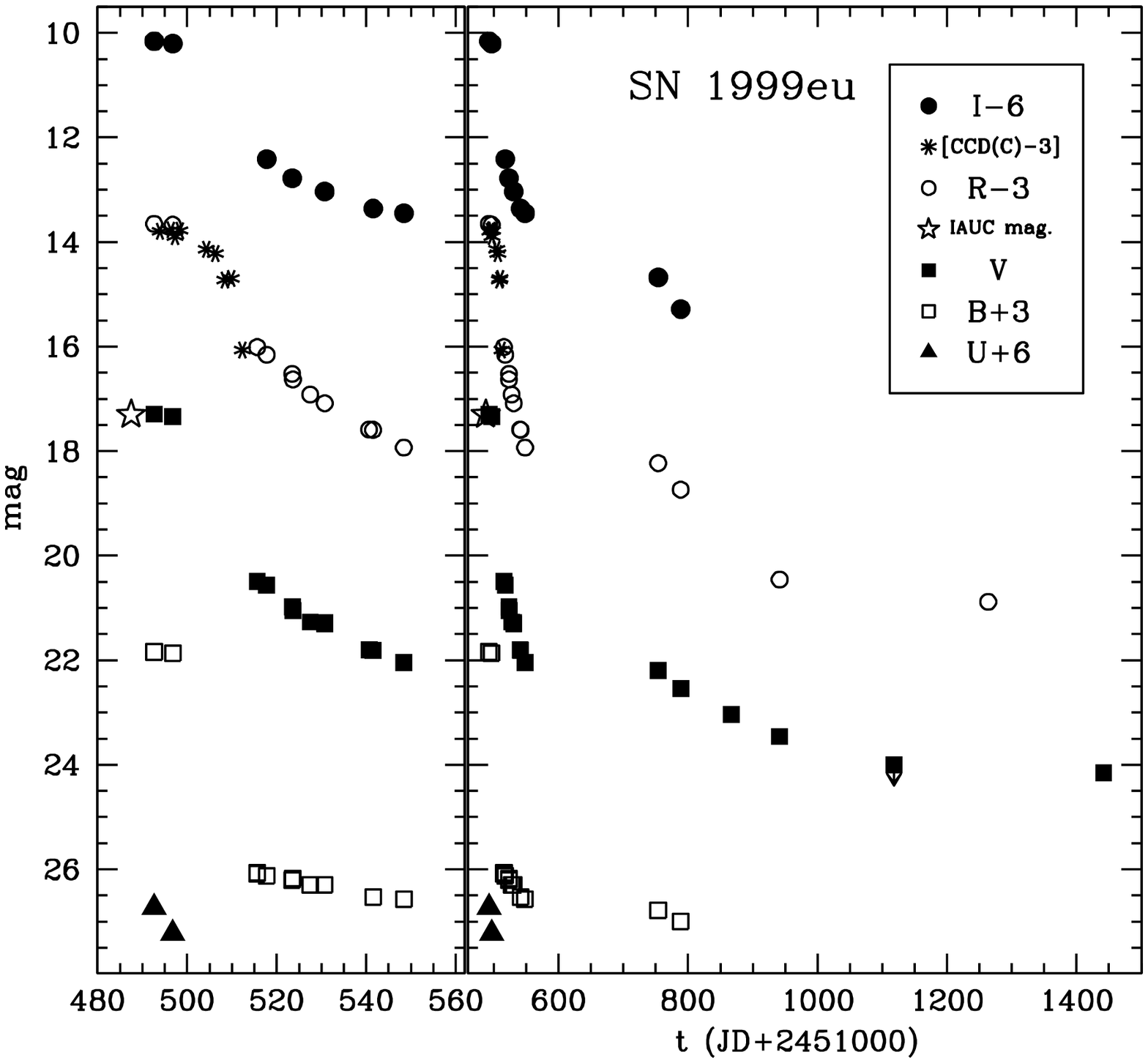}
\caption{BVRI light curves of SN 1999eu. Left: enlargement of the
first two months. Right: complete light curves.  Two points in the U
band are also reported.  The star (plotted with respect to the V
magnitude scale) refers to the discovery unfiltered magnitude (Nakano
\& Aoki, 1999, IAUC 7304).  The asterisks are taken from the web sites
{\sl http://aude.geoman.net/observation/SN/infc99eu.htm} and VSNET
({\sl http://www.kusastro.kyoto-u.ac.jp/vsnet/)}, and are plotted with
respect to the R magnitude scale.} \label{lc_99eu}
\end{figure}

\subsection{SN~1999eu}

Our photometry, covering a period of $\sim$ 950 days, is shown in
Tab. \ref{sn99eu_ph}. Lacking a reference image, the SN magnitudes
were obtained using a PSF--fitting technique.  The errors were
estimated using artificial stars, as described above. In
Tab.\ref{sn99eu_seq} we give the magnitudes of the local sequence
stars (see Fig. \ref{fig_99eu}).  SN~1999eu was observed twice in the
IR with NTT + SOFI. The JHKs magnitudes are also reported in
Tab. \ref{sn99eu_ph}.  From these observations alone it is impossible
to reach a definitive conclusion on the evolution of the light curve
in the IR bands, but they can help in evaluating the bolometric
correction at selected epochs.  Comparison with the IR light curves of
CC--SNe compiled by Mattila \& Meikle (2001) places the earlier JHK
photometry of SN~1999eu about 1.2 magnitudes fainter than the average
values for ``ordinary'' CC--SNe for this phase ($\sim$100~days), and
similar in magnitude to SN~1982R. 
We stress that the last very faint detection could also be due 
to IR background emission.

About 1
week after discovery, the light curves underwent an abrupt fall in
brightness (especially at shorter wavelength bands). About 3~weeks
post--discovery, a slower decline took over. This was observed to
continue until the SN was lost behind the Sun about 9 weeks
post--discovery.  When the SN was recovered $\sim$7 months later, it
was only marginally fainter. The second season light curves
(especially the bolometric curve, cfr. Sect. 5) follow roughly the
$^{56}$Co decay exponential decline rate.  The slopes in the B, V, R
and I passbands are given in Tab. \ref{slope_sn99eu}.  The last two
very faint detections possibly indicate a flattening in the light
curve, but we cannot exclude strong background contamination.
By analogy with the other faint SNe
(see Sect. 6) we believe that SN~1999eu was discovered near the end of
the plateau phase, about 3 months after the explosion.  Unfortunately,
useful constraints on the length of the plateau phase are not
available since the latest prediscovery image was taken about 1 year
before discovery (Nakano and Aoki, 1999).

\begin{table*}
\caption{Optical and infrared photometry of SN 1999eu (JD +2400000).} \label{sn99eu_ph}
\scriptsize
\begin{tabular}{|c|c|c|c|c|c|c|c|c|c|c|} \hline
Date & JD & U & B & V & R & I & J & H & Ks & Ins. \\ \hline \hline  
10/11/99&51492.64 & 20.73 (.08) & 18.84 (.01) & 17.29 (.01) & 16.65 (.01) & 16.16 (.01) & -- & -- & -- & 1 \\  
12/11/99&51494.69 & -- & -- & -- & -- & -- & 15.81 (.01) &  15.61 (.01) & 15.50  (.02) & 7 \\
13/11/99&51496.79 & 21.23 (.06) & 18.87 (.01) & 17.34 (.01) & 16.67 (.01) & 16.21 (.01) & --  & -- & -- & 2 \\ 
02/12/99&51515.66 & -- & 23.06 (.13) & 20.50 (.13) & 19.01 (.04) & -- & -- & -- & -- &4 \\ 
02/12/99&51515.66 & -- & 23.09 (.23) & -- & -- & -- & -- & -- & -- & 4  \\
04/12/99&51517.76 & -- & 23.13 (.33) & 20.56 (.05) & 19.16 (.05) & 18.41 (.18) &  --  & -- & -- &2 \\ 
10/12/99&51523.48 & -- & 23.21 (.19) & 20.98 (.07) & 19.52 (.02) & 18.78 (.12) & --  & -- & -- & 5  \\
11/12/99&51523.64 & -- & 23.17 (.17) & 21.05 (.09) & 19.63 (.02) & -- & -- & -- & -- & 4 \\
15/12/99&51527.50 & -- & 23.29 (.23) & 21.27 (.06) & 19.92 (.02) & -- & -- & -- & -- &5 \\
18/12/99&51530.73 & -- & 23.30 (.39) & 21.28 (.12) & 20.08 (.04) & 19.04 (.03) & -- & -- & -- &2\\  
18/12/99&51530.73 & -- & -- & 21.31 (.13) & -- & -- & -- & -- & -- & 2   \\
28/12/99&51540.65 & -- & -- & 21.80 (.61) & 20.59 (.21) & -- & -- & -- & -- &4\\  
28/12/99&51540.65 & -- & $\ge$22.68 (.28) & -- & -- & -- & -- & -- & -- &4 \\
29/12/99&51541.53 & -- & 23.53 (.17) & 21.81 (.12) & 20.59 (.02) & 19.36 (.01) & -- & -- & -- &1 \\ 
04/01/00&51548.39 & -- & 23.58 (.18) & 22.05 (.08) & 20.93 (.03) & 19.45 (.19) & -- & -- & -- &5 \\ 
27/07/00&51753.92 & -- & 23.78 (.17) & 22.20 (.10) & 21.23 (.08) & 20.68 (.03) & -- & -- & -- & 3 \\
31/08/00&51788.85 & -- & 24.00 (.40) & 22.54 (.25) & 21.74 (.16) & 21.28 (.11) & -- & -- & -- & 2 \\
10/11/00&51858.73 & -- & -- & -- & -- & -- & 21.93 (.43) & 20.62 (.34) & 20.50 (.30) & 7\\
19/11/00&51866.71 & -- & -- & 23.04 (.10) & -- & -- & -- & -- & -- &1  \\
01/02/01&51941.54 & -- & -- & 23.46 (.13) & 23.46 (.10)  & -- & -- & -- & -- &2\\ 
27/07/01&52117.87 & -- & -- & $\ge$24.0  & --  & -- & -- & -- & -- &6\\  
20/12/01&52263.68 & -- & -- & -- & 23.88 (.17) & -- & -- & -- & -- &6 \\ 
23/06/02&52441.87 & -- & -- & 24.15 (.40)& --  & -- & -- & -- & -- &6\\  \hline 
\end{tabular}

1 = ESO Dan1.54 + DFOSC; 2 = ESO 3.6m + EFOSC2; 3 = ESO NTT + EMMI; \\
4 = ESO 2.2m + WFI + chip7, 5 = TNG + OIG; 6 = VLT + FORS1 ; 7 = NTT + SOFI\\
\end{table*}

\begin{table*}
\caption{Magnitudes of the sequence stars in the field of NGC
1097. The numbers in brackets are the r.m.s. of the available
measurements. If no error is reported, only a single estimate is
available. We estimated the errors in J, H, Ks as the deviation from
the average of two available measurements.} \label{sn99eu_seq}
\scriptsize
\begin{tabular}{|c|c|c|c|c|c|c|c|c|} \hline
Star & U & B & V & R & I & J & H & Ks \\ \hline \hline  
1 & 21.78 (--) & 21.13 (0.01) & 19.95 (0.02) & 19.20 (0.01) & 18.44 (0.01) & 17.31 (0.10) & 16.97 (0.01) & 16.94 (0.01)\\
2 & 18.14 (0.01) & 17.68 (0.02) & 16.77 (0.01) & 16.19 (0.03) & 15.72 (0.02) & 15.16 (0.02) & 14.63 (0.02) & 14.67 (0.02)\\
3 & 19.13 (--) & 18.41 (0.01) & 16.64 (0.01) & 15.56 (0.03) & 14.01 (0.06) & -- & -- & --\\
4 & 20.24 (--) & 19.26 (0.02) & 18.00 (0.01) & 17.14 (0.01) & 16.41 (0.01) & 15.55 (0.10) & 14.91 (0.04) & 14.84 (0.03)\\
5 & -- & 23.31 (0.01) & 21.43 (0.01) & 20.41 (0.02) & 19.40 (0.02) & 18.12 (0.06) & 17.35 (0.12) & 16.93 (0.05)\\ 
6 & 19.45 (0.01) & 18.49 (0.02) & 17.58 (0.02) & 16.94 (0.03) &  16.38 (0.01) & -- & -- & --\\
7 & 17.23 (0.03) & 16.95 (0.03) & 16.23 (0.01) & 15.77 (0.03) & 15.38 (0.01) & 14.90 (0.01) & 14.51 (0.02) & 14.51 (0.06)\\ 
8 & 22.85: (--) & 23.04 (0.05) & 21.22 (0.06) & 20.45 (0.04) &  20.08 (0.01) & 18.86 (0.09) & 18.03 (0.02) & 17.78 (0.11) \\
9 & 21.06 (--) & 21.26 (0.02) & 20.63 (0.01) & 20.23 (0.02) & 19.80 (0.03) & 19.34 (0.02) & -- & -- \\
10 & 22.29 (--) & 21.57 (0.02) & 20.59 (0.03) & 19.97 (0.02) & 19.52 (0.06) & 18.61 (0.11) & -- & 17.99 (0.03)\\ \hline
\end{tabular}
\end{table*}

\begin{table}
\caption{Slopes of the light curves of SN 1999eu in the B, V, R, I
bands (in mag/100$^{d}$).  The phase is with respect to the day of
explosion (JD = 2451394).} \label{slope_sn99eu} \footnotesize
\begin{center}
\begin{tabular}{|c|c|c|c|c|c|} \hline
band & 95--125 & 120--160 & 150--365 & 355--550 & 540--1050\\ \hline \hline  
$\gamma_B$ & 19.77 & 1.60 & 0.11 & 0.61 & --\\
$\gamma_V$ & 14.31 & 4.78 & 0.07 & 0.66 & 0.14\\
$\gamma_R$ & 10.05 & 5.68 & 0.14 & 1.17 & 0.13\\
$\gamma_I$ &  9.48 & 3.30 & 0.60 & 1.74 & --\\ \hline
\end{tabular}
\end{center}
\end{table}

\subsection{SN~1994N}

\begin{figure}
\includegraphics[width=8.05cm]{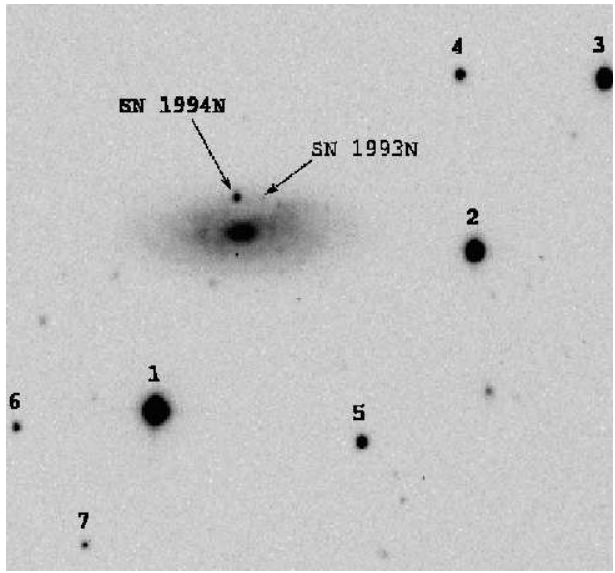}
\caption{SN 1994N and the type IIn SN 1993N in UGC 5695 (R band image
taken with the ESO 3.6m telescope + EFOSC1 on 1994 May 9).}\label{sn94N_fig}
\end{figure}

\begin{figure}
\includegraphics[width=8.5cm]{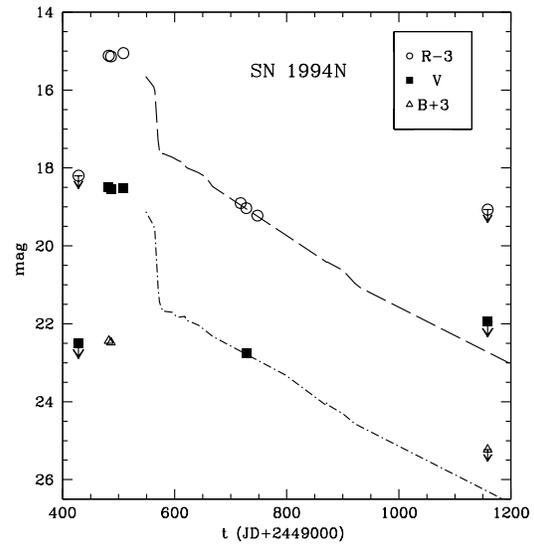}
\caption{B, V, R band light curves of SN 1994N. The dashed lines
represent the V and R band light curves of SN 1997D shifted in
magnitude and in time to match the points of SN 1994N, plotted 
to guide the eye.}\label{sn94N_lc}
\end{figure}

\begin{table*}
\caption{Photometry of SN 1994N (JD +2400000).} \label{sn94N_ph}
\scriptsize
\begin{tabular}{|c|c|c|c|c|c|c|c|} \hline
Date & JD & U & B & V & R & I & Instrument \\ \hline \hline  
17/03/94 & 49428.5  & -- & -- & $\ge$22.50  & $\ge$21.20  & -- & 1 \\ 
09/05/94 & 49482.48 & -- &  19.44 (0.04) & 18.50 (0.02) & 18.12 (0.02) & -- & 2 \\ 
13/05/94 & 49486.48 & 20.44 (0.25) & 19.48 (0.01) & 18.55 (0.01) & 18.14 (0.01) & 17.95 (0.02) & 3 \\
05/06/94 & 49508.52 & -- & -- & 18.52 (0.05) & 18.05 (0.04) & -- &  2 \\
31/12/94 & 49717.76 & -- & -- & -- & 21.90 (0.15) & -- &  3 \\
10/01/95 & 49727.80 & -- & -- & 22.77 (0.03) & 22.03 (0.04) & -- &  4 \\
30/01/95 & 49747.74 & -- & -- & -- & 22.22 (0.08) & -- &  1 \\
16/03/96 & 50158.5  & -- & $\ge$22.23  & $\ge$22.95  & $\ge$22.07  & -- & 2 \\ \hline
\end{tabular}

1 = NTT + EMMI; 2 ESO 3.6m + EFOSC1; 3 = ESO 2.2m + EFOSC2; 4 = NTT + SUSI 
\end{table*}

\begin{table*}
\caption{Magnitudes of the sequence stars in the field of UGC
5695. The numbers in brackets are the r.m.s. of the available
measurements. Uncalibrated U and I magnitudes of sequence stars
obtained on 1994 May 13 are also reported.} \label{sn94N_seq}
\scriptsize
\begin{tabular}{|c|c|c|c|c|c|} \hline
Star & U & B & V & R & I \\ \hline \hline  
2 & -- & -- & 14.54 (0.01) & 14.16 (0.01) & -- \\
3 & -- & -- & 14.93 (0.03) & 14.49 (0.01) & -- \\
4 & 18.19 (--) & 18.21 (0.02) & 17.47 (0.01) & 17.04 (0.02) & 16.77 (--) \\
5 & 17.33 (--) & 17.60 (0.02) & 16.98 (0.01) & 16.67 (0.01) & 16.40 (--) \\ 
6 & -- & 20.06 (0.02) & 18.85 (0.01) & 18.21 (0.01) & 17.74 (--) \\
7 & -- & -- & 20.24 (0.01) & 18.91 (0.04) & 17.32 (--) \\ \hline
\end{tabular}
\end{table*}

The photometry of SN 1994N is sparse (see Tab. \ref{sn94N_ph}). It was
obtained using ESO--La Silla telescopes, often under poor weather
conditions.  The late--time photometry was performed using the template
subtraction technique. Again, the errors were estimated using
artificial stars. As standard stars we adopted the local sequence in
the field of UGC~5695 calibrated during some photometric nights in
1993 (during the follow--up of SN~1993N). The magnitudes of the
sequence stars are given in Tab. \ref{sn94N_seq}.  The light curve
shows that this SN was discovered during the plateau phase.  The
prediscovery upper limits constrain the explosion epoch to no earlier
than $\sim$7~weeks before discovery (see Fig. \ref{sn94N_lc}).  We
measure the slope during the plateau and find respectively $\gamma_V$ $\approx$ 0
and $\gamma_R$ $\approx -$0.3 mag/100$^{d}$, although,
given that there are only a few points in the V and R band light
curves, we regard this measurement as tentative. 
At late times, the observations indicate $\gamma_V$ $\approx$ 1
mag/100$^{d}$, although the photometric errors are sometimes quite large.

\subsection{SN~2001dc}

\begin{figure}
\includegraphics[width=7.85cm]{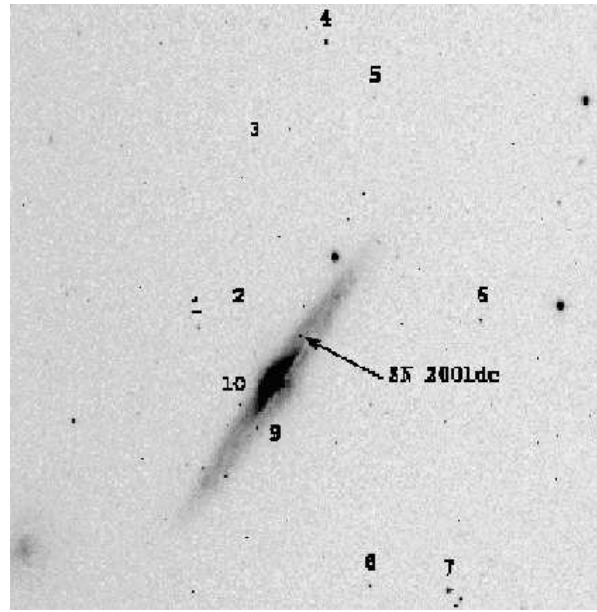}
\caption{SN 2001dc in NGC 5777 (R band image taken
with JKT + IAG ccd on 2001 July 13).} \label{sn01dc_field}
\end{figure}

\begin{figure}
\includegraphics[width=8.65cm]{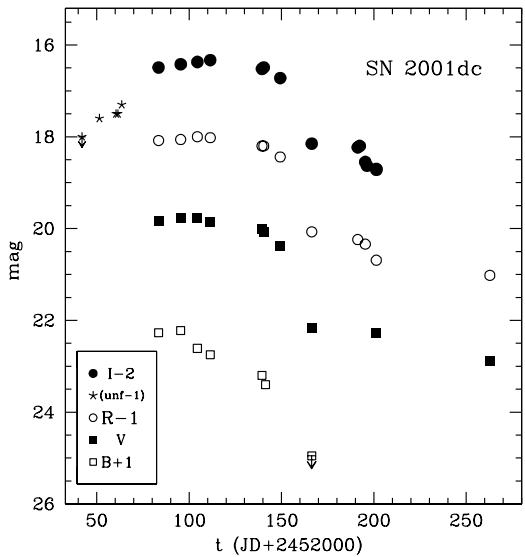}
\caption{BVRI light curves of SN 2001dc.  The unfiltered prediscovery
limit and magnitudes (asterisks) are
shown with respect to the R magnitude scale (Hurst et al., 2001). The
magnitude at phases later than $\sim$ 100 days are affected by large
errors.}
\label{sn01dc_lc}
\end{figure}

\begin{table*}
\caption{Photometry of SN 2001dc (JD +2400000).} \label{sn01dc_ph}
\scriptsize
\begin{tabular}{|c|c|c|c|c|c|c|c|} \hline
Date & JD & U & B & V & R & I & Instrument \\ \hline \hline  
22/06/01 & 52083.43 & -- & 21.27 (0.06) & 19.84 (0.03) & 19.08 (0.01) & 18.49 (0.01) & 1 \\
04/07/01 & 52095.42 & -- & 21.22 (0.09) & 19.78 (0.03) & 19.06 (0.02) & 18.42 (0.02) & 1 \\
13/07/01 & 52104.46 & -- & 21.61 (0.08) & 19.78 (0.04) & 19.00 (0.03) & 18.37 (0.02) & 1 \\
20/07/01 & 52111.44 & 23.65 (0.44) & 21.75 (0.04) & 19.86 (0.04) & 19.02 (0.02) & 18.33 (0.02) & 2 \\
17/08/01 & 52139.42 & -- & 22.20 (0.26) & 20.00 (0.10) & 19.20 (0.07) & 18.52 (0.04) &  3 \\
18/08/01 & 52140.39 & -- &  -- & 20.08 (0.06) & 19.20 (0.05) & 18.49 (0.03) & 4  \\
19/08/01 & 52141.36 & -- & 22.40 (0.34) & -- & -- & -- & 4 \\
27/08/01 & 52149.32 & -- & -- & 20.38 (0.19) & 19.44 (0.15) & 18.72 (0.05) & 4 \\
13/09/01 & 52166.36 & -- & $\geq$23.95 & -- & -- & -- & 3 \\
13/09/01 & 52166.36 & -- & -- & 22.16 (0.30) &  21.07 (0.13) & 20.15 (0.08) & 3 \\
08/10/01 & 52191.26 & -- & -- & -- & 21.24 (0.27) & 21.23 (0.37) & 4 \\
09/10/01 & 52192.33 & -- & -- & -- & --  & 20.20 (0.50) & 4 \\ 
12/10/01 & 52195.33 & -- & -- & -- & 21.34 (0.30) & 20.55 (0.30) & 4 \\ 
13/10/01 & 52196.26 & -- & -- & -- & --  & 20.63 (0.15) & 4 \\ 
18/10/01 & 52201.33 & -- & -- & 22.27 (0.65) & 21.69 (0.30) & 20.70 (0.27) & 3 \\  
18/10/01 & 52201.35 & -- & -- & -- & --  & 20.72 (0.27) & 3 \\ 
19/12/01 & 52262.76 & -- & -- & 22.89 (0.30) & 22.02 (0.32) & -- & 3\\ \hline 
\end{tabular}

1 = JKT + JAG; 2 = TNG + OIG; 3 = TNG + Dolores; 4 = Asi1.82m + AFOSC \\
\end{table*}

\begin{table*}
\caption{Magnitudes of the sequence stars in the field of NGC
5777. The numbers in brackets are the r.m.s. of the available
measurements.  Uncalibrated U magnitudes of sequence stars obtained on
2001 July 20 are also reported.} \label{sn01dc_seq} \scriptsize
\begin{tabular}{|c|c|c|c|c|c|} \hline
Star & U & B & V & R & I \\ \hline \hline  
1 & 19.72 (--) & 19.68 (0.01) & 18.88 (0.01) & 18.40 (0.01) & 17.94 (0.02) \\
2 & 20.45 (--) & 20.97 (0.02) & 20.56 (0.01) & 20.28 (0.01) & 20.06 (0.01) \\
3 & -- & 20.35 (0.01) & 19.92 (0.01) & 19.64 (0.01) & 19.28 (0.01) \\
4 & -- & 19.24 (0.01) & 18.70 (0.01) & 18.37 (0.01) & 18.04 (0.01) \\
5 & -- & 20.87 (0.02) & 20.42 (0.02) & 20.09 (0.01) & 19.88 (0.02) \\
6 & 21.60 (--) & 20.77 (0.01) & 19.26 (0.01) & 18.33 (0.01) & 17.38 (0.01) \\
7 & 20.51 (--) & 19.71 (0.03) & 18.51 (0.01) & 17.79 (0.01) & 17.10 (0.02) \\
8 & 19.22 (--) & 19.21 (0.01) & 18.33 (0.02) & 17.85 (0.01) & 17.40 (0.01) \\
9 & 19.75 (--) & 19.43 (0.01) & 18.57 (0.01) & 18.02 (0.01) & 17.56 (0.02) \\
10 & -- & 22.16 (0.02) & 20.75 (0.02) & 19.83 (0.03) & 18.87 (0.04) \\ \hline
\end{tabular}
\end{table*}

Our photometry measurements of SN~2001dc, lacking template images,
were obtained using the PSF fitting technique.  The magnitudes are
reported in Tab.  \ref{sn01dc_ph}, while the light curves are shown in
Fig. \ref{sn01dc_lc}. The large errors at late times (estimated with
the artificial stars method) are caused by the complex background upon
which the SN is projected.  Tab. \ref{sn01dc_seq} gives the magnitudes
of the local sequence stars (Fig. \ref{sn01dc_field}).  In the
following, we adopt 2001 May 17 (JD = 2452047 $\pm$5) as the explosion 
epoch, this being
intermediate between the upper limit of May 12 and the first (prediscovery) 
detection on May 21.98. \\ 
The light curve shows a plateau persisting to 90--100 days
post--explosion. The slope of the light curves for three
representative epochs are reported in Tab. \ref{slope_sn01dc}. Despite
the large errors due to the faintness of the SN and its position, the
slopes in V and R during $\sim$115--220 days are rather close to the
luminosity decline rate of $^{56}$Co.\\

\begin{table}
\caption{Slope of the light curve of SN 2001dc in the B, V, R, I bands
(in mag/100$^{d}$). The phase is relative to JD = 2452047.}
\label{slope_sn01dc} \footnotesize
\begin{center}
\begin{tabular}{|c|c|c|c|} \hline
band & 35--70 & 90-120 & 115--220 \\ \hline \hline  
$\gamma_B$ & 1.88 & -- & -- \\
$\gamma_V$ & 0.02 & 7.97 & 0.79 \\
$\gamma_R$ & -0.27 & 8.00 & 0.90 \\
$\gamma_I$ & -0.57 & 6.17 & 1.59 \\ \hline
\end{tabular}
\end{center}
\end{table}

\section{COLOR EVOLUTION, ABSOLUTE LUMINOSITY AND BOLOMETRIC LIGHT CURVES}

\begin{figure}
\includegraphics[width=9.15cm]{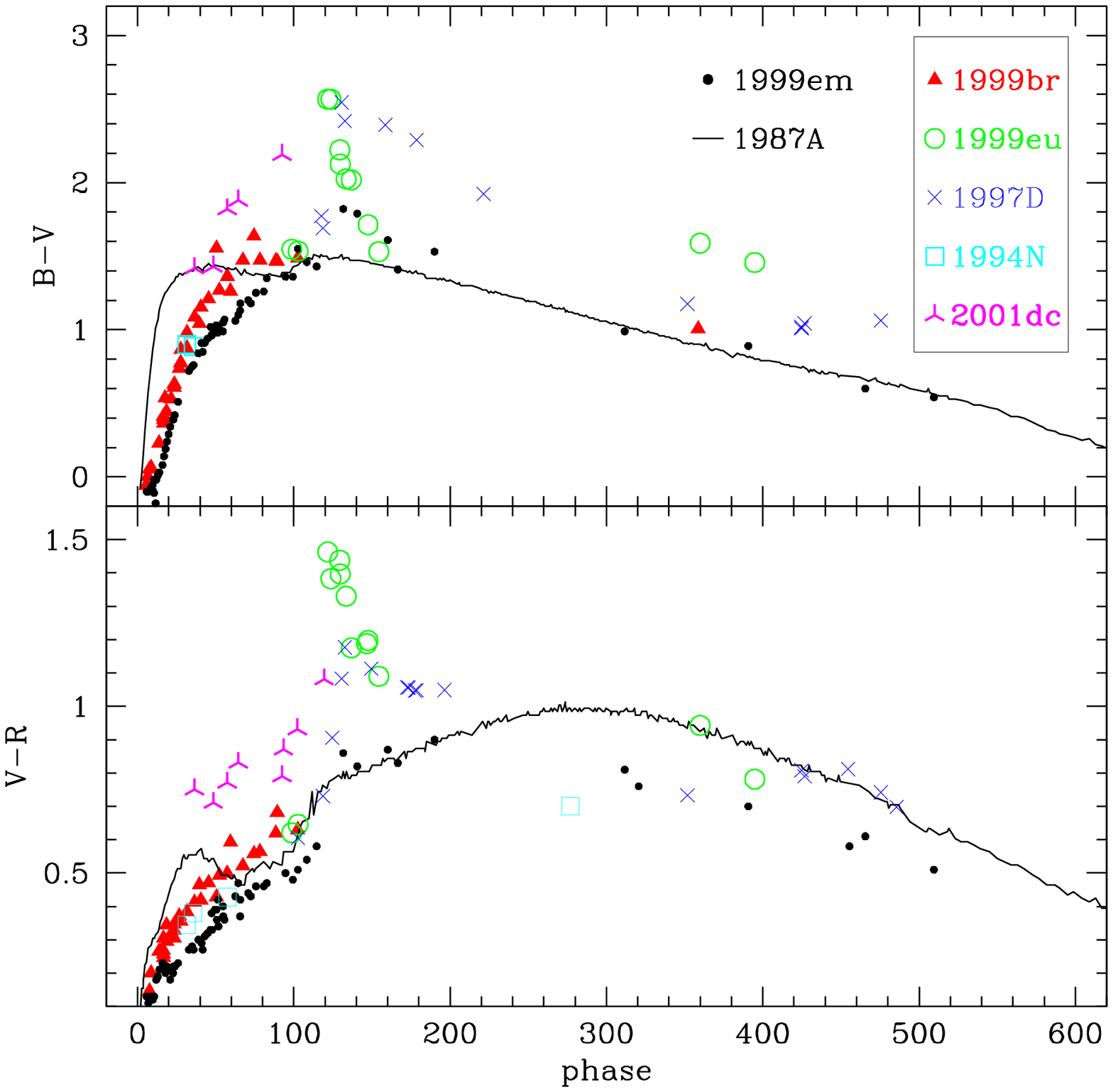}
\caption{Evolution of (B-V) and (V-R) colours of the sample SNe,
compared with those of SN~1987A and SN~1999em. The SN~2001dc colours are not plotted beyond
$\sim$120 days because of their large uncertainty.  Corrections for
galactic extinction only have been applied to the sample CC--SNe.  The
SN 1987A and SN 1999em colours have been corrected for the total extinction 
to these events, using respectively A$_V$ = 0.6 and A$_V$ = 0.31 .}\label{color}
\includegraphics[width=9.15cm]{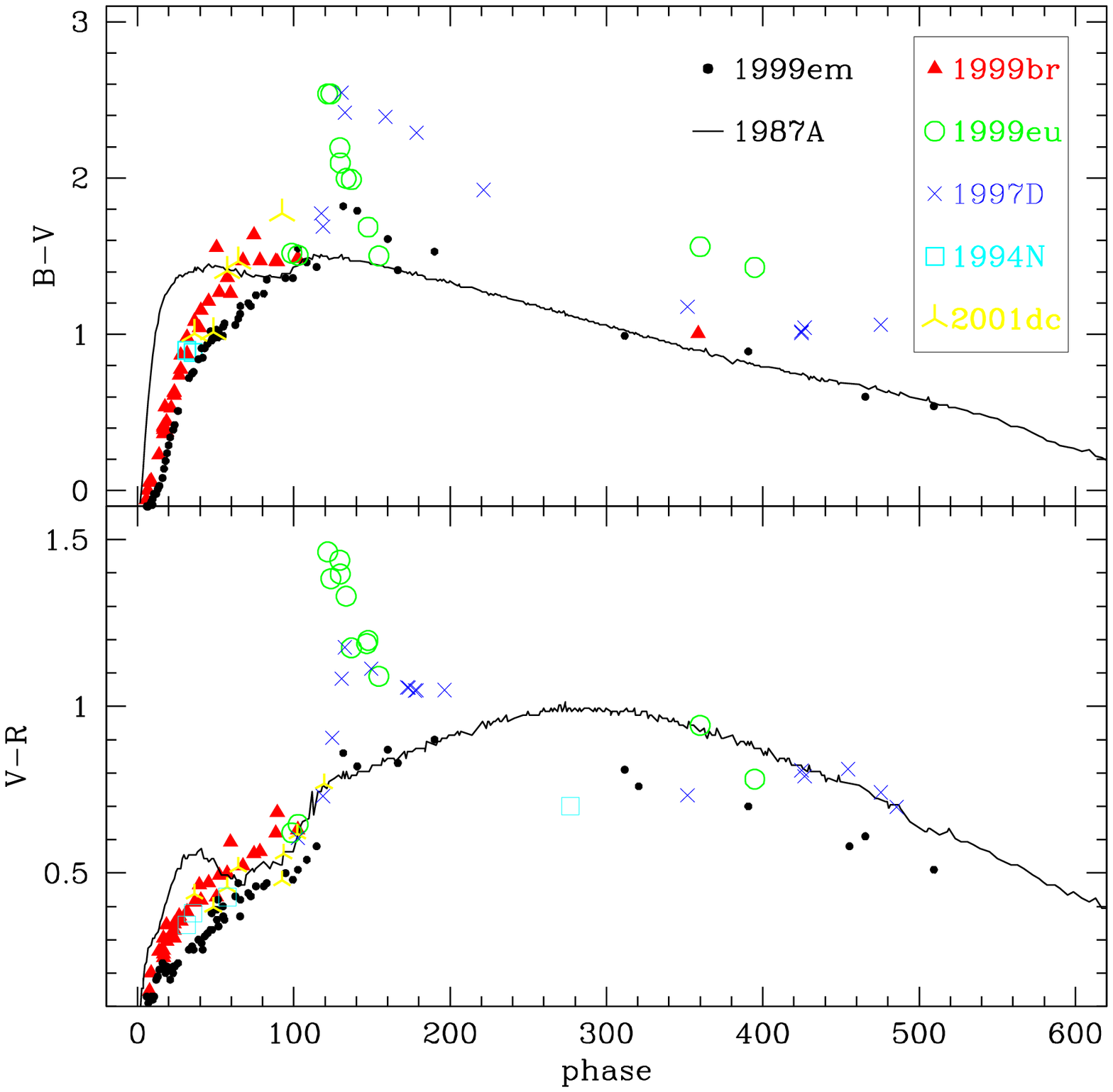}
\caption{Same as Fig. \ref{color}, but correcting for the total
extinction towards SN~2001dc, using E(B-V) = 0.42, E(V-R) = 0.33
magnitudes.}\label{color2}
\end{figure}

In order to establish the intrinsic luminosity of our sample of
supernovae, it is essential to establish their distances, phases and
the extent to which they are subject to extinction.  We have already
estimated the distance moduli (Section 2) and these are listed in
Tab.~1.   \\

As explained in Section 4, prediscovery limits for SNe~1999br and
2001dc allow us to pin down their explosion epochs to within a few
days and hence establish clearly the phase of these events at any
given date. In addition, below we shall show that the similarity of
the spectrum of SN~1999br at a phase of $\sim$100 days to the earliest
available spectra of SNe 1997D and 1999eu allows us to fix the phases
of these two events. Likewise, we determined the phase of SN~1994N
through the similarity of its earliest spectrum to a spectrum of
SN~1999br at $\sim$30~days. Thus, we deduce that the phases of our
five CC--SNe at discovery were: SN 1994N: 31.5 days, SN 1997D: 101.7 days, 
SN 1999br: 2.9 days, SN 1999eu: 93.5 days, SN 2001dc: 4.5 days.\\

For extinction, we have argued that for SNe 1994N, 1997D, 1999br and 1999eu we
need only to consider the effect of the Milky Way ISM.
However, to estimate the extinction towards SN~2001dc, we must first
consider the colours and colour evolution of the five CC--SNe.  This
is also valuable as an additional mean of testing the degree of
homogeneity of the sample. The (B--V) and (V--R) colours are shown as a
function of phase and compared with those of the peculiar SN~1987A and 
the ``normal'' type II--P SN~1999em in 
Fig. \ref{color}.  Corrections for galactic extinction only have been 
applied to the sample CC--SNe.  The SN 1987A and SN 1999em
colours have been corrected for the total extinction to these events,
using respectively A$_V$ = 0.6 \cite{west87} and A$_V$ = 0.31 \cite{baro00}.  
For SN~2001dc, the similarity of its colour {\it evolution} and its spectra to those of SN~1999br at similar epochs
during the plateau phase, plus its location close to a dusty region in
NGC 5777, leads us to argue that its redder colours (relative to
SN~1999br) are due to strong extinction in the host galaxy (see
Fig. \ref{sn01dc_field}) as well as in the Milky Way.  Therefore, in
Fig. \ref{color2}, we show the same data, but with the SN~2001dc
colours dereddened to match those of SN~1999br. From this, we deduce
the total reddening to SN~2001dc to be E(B-V) = 0.42 and
E(V-R) = 0.33, corresponding to A$_{B}$ = 1.7. The adopted
extinction values for all the SNe are shown in Table~1.\\

\begin{figure*}
\includegraphics[width=10.9cm,height=14.0cm,angle=270]{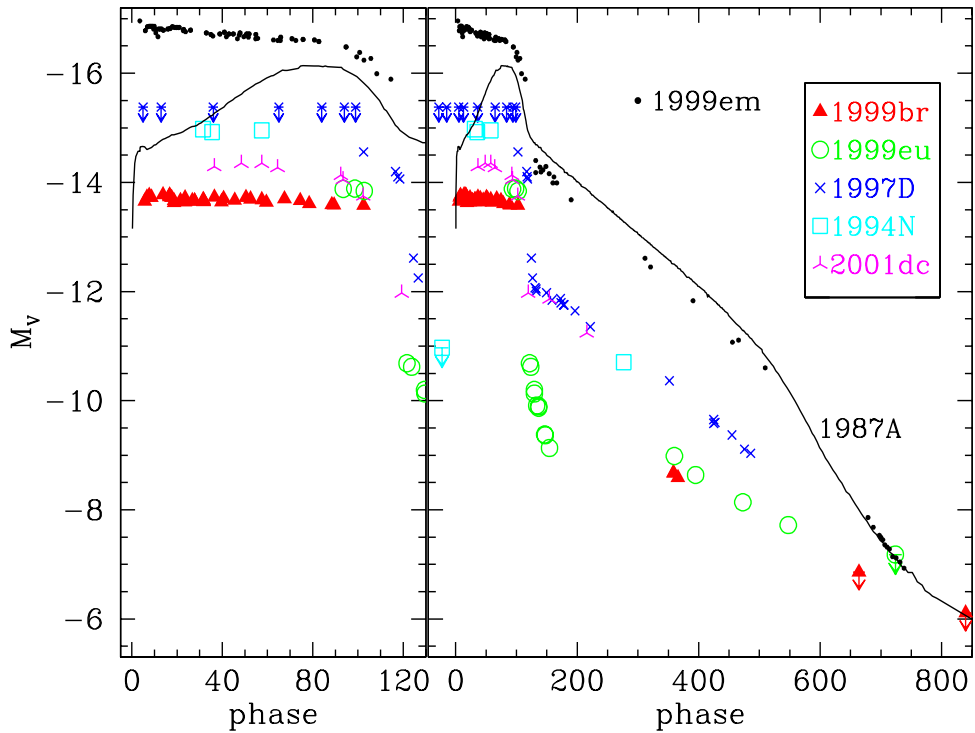}
\caption{V band absolute light curves of the sample CC--SNe discussed
in this paper, SN~1987A and SN~1999em. By comparison of spectra (Sect. 6), we
conclude that the explosion epochs of SNe 1997D, 1999eu and 1994N
were, respectively 101.7, 93.5 and 31.5 days before discovery.
Prediscovery upper limits (unfiltered) are also shown along with some
late--time upper limits.  The SN~1987A light curve (see Patat et al.,
1994 and references therein) are corrected for total extinction A$_V$
= 0.6, that of SN 1999em (Hamuy et al., 2001; Leonard et al., 2002a; 
Elmhamdi et al., 2003; Leonard et al., 2003) for A$_V$
= 0.31.  The SN~2001dc photometry has been corrected for a total
estimated extinction of A$_V$ = 1.28 (Sect. 5).  The host galaxy
extinction for the other 4 SNe is probably negligible, and so the
photometry for these SNe has been corrected for Galactic extinction
only (Schlegel et al. 1998).}\label{abs_V}
\end{figure*}

\begin{figure}
\includegraphics[width=9cm]{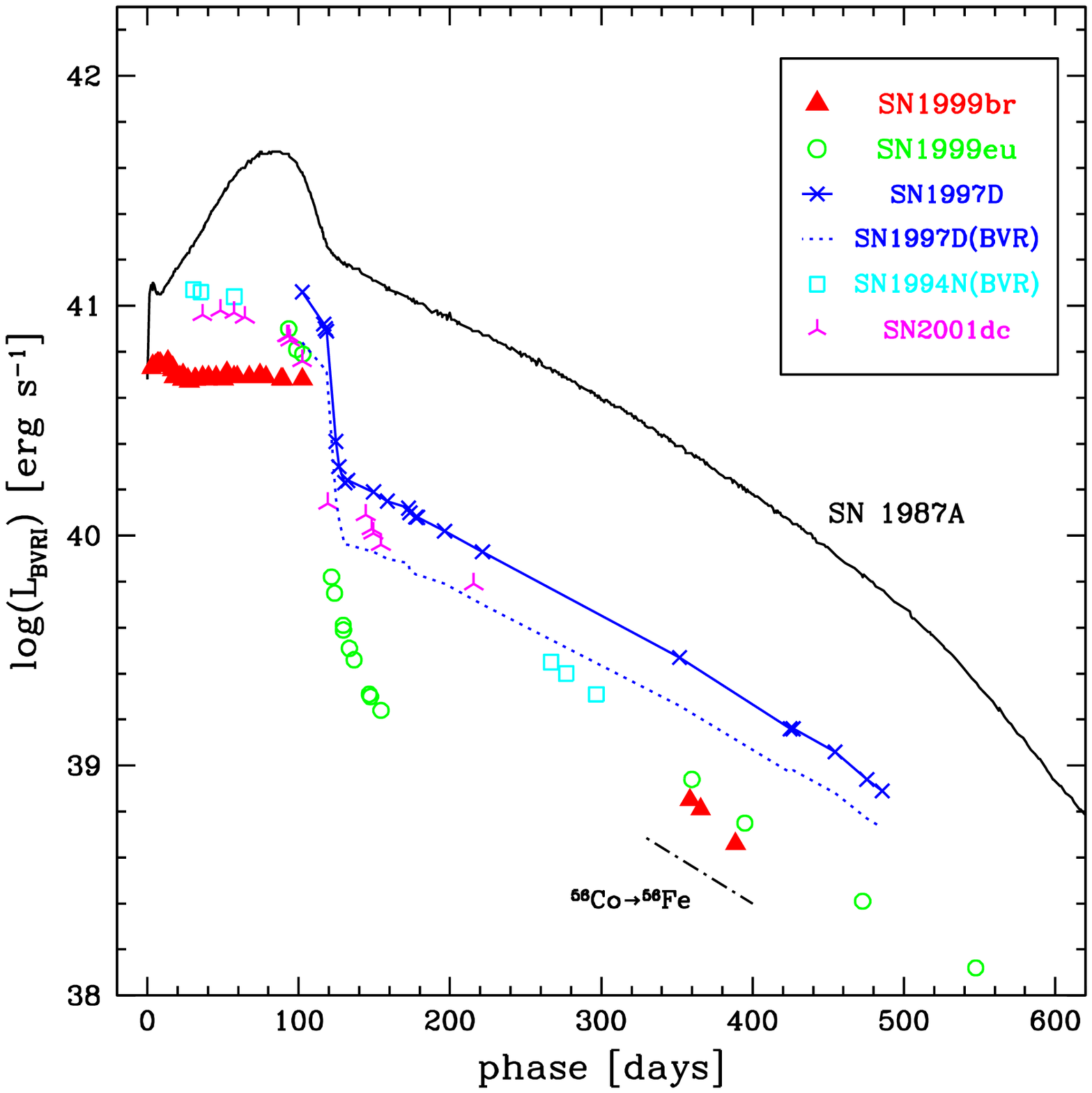}
\caption{Comparison of the pseudo--bolometric {\sl ``OIR''} light
curves of the sample CC--SNe.  The data of SN~1997D are from Turatto
et al. (1998) and Benetti et al. (2001). The bolometric
light curve of SN~1987A (Patat et al., 1994 and reference therein) is
also shown for comparison. For SN 1994N, the integration is limited to
the BVR bands.  To illustrate the effect of this limitation, we show
also the equivalent BVR light curve of SN 1997D (dotted
line).}\label{bolom}
\end{figure}

Inspection of Fig. \ref{color2} shows that, during the first 100 days,
the data are consistent with there being a similar colour evolution
for the 5 sample SNe.  During the plateau phase the (B--V) colour reddens
for the first $\sim$60~days reaching (B--V) $\sim$ 1.5. It then remains
at about this value for the next 60~days.  Beyond $\sim$100~days, good
photometric coverage is available for only 2 of the 5 sample SNe
viz. for SNe 1997D and 1999eu. At $\sim$120~days, the (B--V) colour of
these two events suddenly reddens further to (B--V) $\sim$ 2.5.  This
coincides with the epoch of the steep post--plateau decline
(cf. Figs. 5 and 12).  After this, the (B--V) colour of these two SNe
become bluer, with SN~1999eu showing a particularly rapid change.  The
plot of (V--R) versus time shows a rather similar behaviour.  
SNe~1987A and 1999em show a somewhat different behaviour, and in particular 
does not show a rapid change at 120--160~days.\\

Having established the distances, phases and extinctions for the five
CC--SNe, we derived M$_V$ light curves for these events.  These light
curves are plotted in Fig. \ref{abs_V} and compared with that of SNe 1987A
and 1999em.
It can be immediately seen that for all epochs at which photometry is
available, i.e. during both the plateau and nebular phases, the 5
sample SNe are fainter than both SNe 1987A and 1999em.  In the case of SN~1999br, this
includes epochs as early as just $\sim$ 1 week after explosion.  Given
that SN 1987A itself was unusually under--luminous, we conclude that
our CC--SNe, at least when observed, were all exceptionally faint. For
example, during the plateau phase, SN 1999br had a magnitude of just
M$_{V} \approx -$13.76, while for SN 2001dc M$_V \approx -$14.29.  We
conclude that, in spite of the incomplete coverage of individual light
curves, it is likely that the five CC--SNe considered here were all
exceptionally subluminous throughout both the plateau phase and the
later radioactive tail.

The pseudo--bolometric {\sl ``OIR''} light curves 
(shown in Fig. \ref{bolom}) were computed integrating the emitted fluxes 
in the B, V, R and I passbands at the epochs for which measurements were 
available, and interpolating between points adjacent in time when a 
measurement was missing. 
In the case of SN 1994N only a single point was available in the I band and
the integration had to be restricted to the BVR bands.  To illustrate
the effect of this limitation, we show also the equivalent BVR light
curve of SN 1997D as a dotted line. 
The 5 SNe are fainter than SN~1987A at any epoch: i.e. the plateau luminosity of 
SN 1999br (the best monitored event) is L$_{BVRI}$ $\approx$ 5 $\times$ 
10$^{40}$ erg s$^{-1}$, a factor of 10 times lower than luminosity of SN~1987A at the 
epoch of the broad maximum.
The earlier light curves of SNe
1994N and 2001dc show similar magnitudes and evolution, while
SN~1999br is about a factor 2 times fainter. Between $\sim$100
and 200~days, SNe~1997D, 1999eu and 2001dc show somewhat different
magnitudes and/or evolution. SN~1999eu shows a particularly dramatic
drop in luminosity during this time.  Between days~270 and 550, the
magnitudes and decline rates of SNe~1999br and 1999eu are roughly
consistent with their being powered at this time by a small amount of
$^{56}$Ni (see also Sect. 7).

\section{Spectroscopy}
The SNe discussed in this paper have also been observed
spectroscopically, and despite their faintness, later coverage extends
into the nebular phase for some events.  In this section we present
spectra for SNe~1994N, 1999br, 1999eu and 2001dc.  We discuss these
spectra together with those of SN~1997D (Benetti et al., 2001). In
particular, we shall illustrate the use of spectral similarities to
determine the phases of some of the CC--SNe.
\subsection{Individual Properties}
    	   
\begin{figure*}
\includegraphics[width=11.5cm,angle=270]{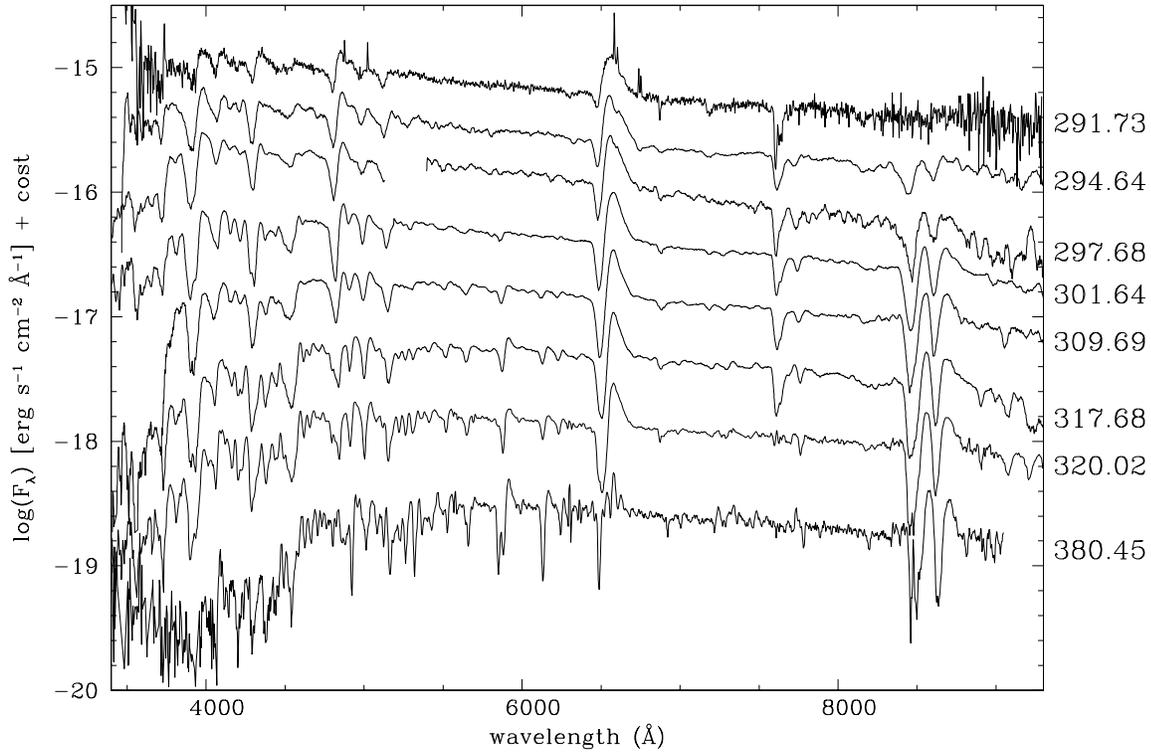}
\caption{Spectral evolution of SN 1999br. The numbers on the right are
JD +2451000. The sequence of the first 7 spectra covers a period of
about one month from discovery, while the last spectrum (1999 July 20,
JD = 2451380.45) is taken 2 months later, at the end of the plateau.} \label{spec_ev_99br}
\end{figure*}

\begin{figure}
\includegraphics[width=9cm]{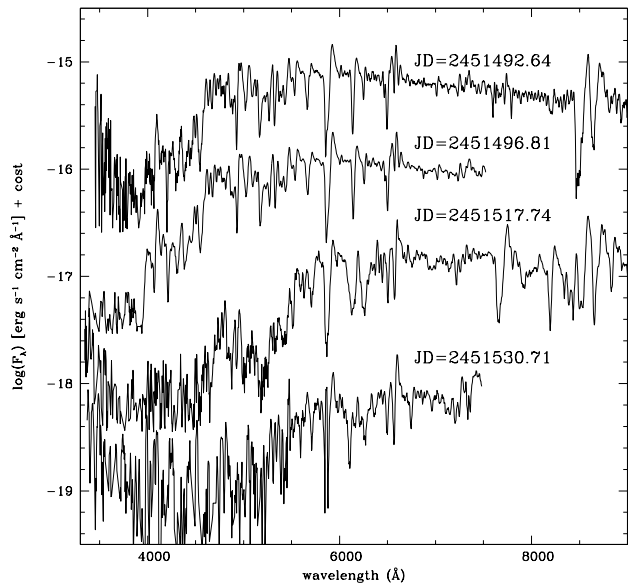}
\caption{Spectral evolution of SN 1999eu.  The spectrum of 1999
December 18 (JD = 2451530.71) is smoothed because of the poor signal
to noise ratio.} \label{spec_ev_99eu}
\end{figure}

\begin{figure}
\includegraphics[width=9cm]{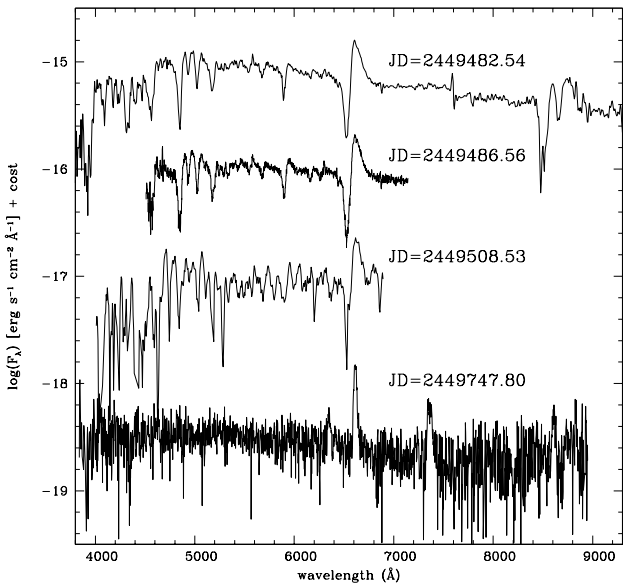}
\caption{Spectral evolution of SN 1994N.  The spectrum of 1994 June 5
(JD = 2449508.53) is smoothed because of the poor signal to noise
ratio.}\label{spec_ev_94N}
\end{figure}

\begin{figure}
\includegraphics[width=9cm]{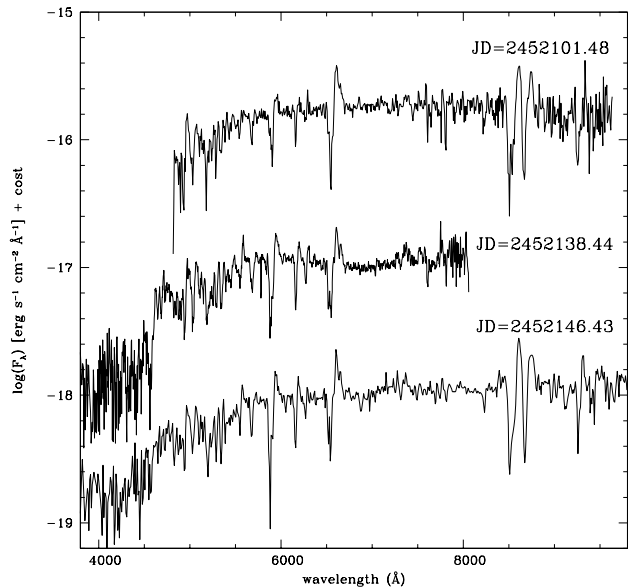}
\caption{Spectral evolution of SN 2001dc. The spectra have not been
corrected for reddening.}
\label{spec_ev_01dc}
\end{figure}
 
\begin{figure}
\includegraphics[width=9cm]{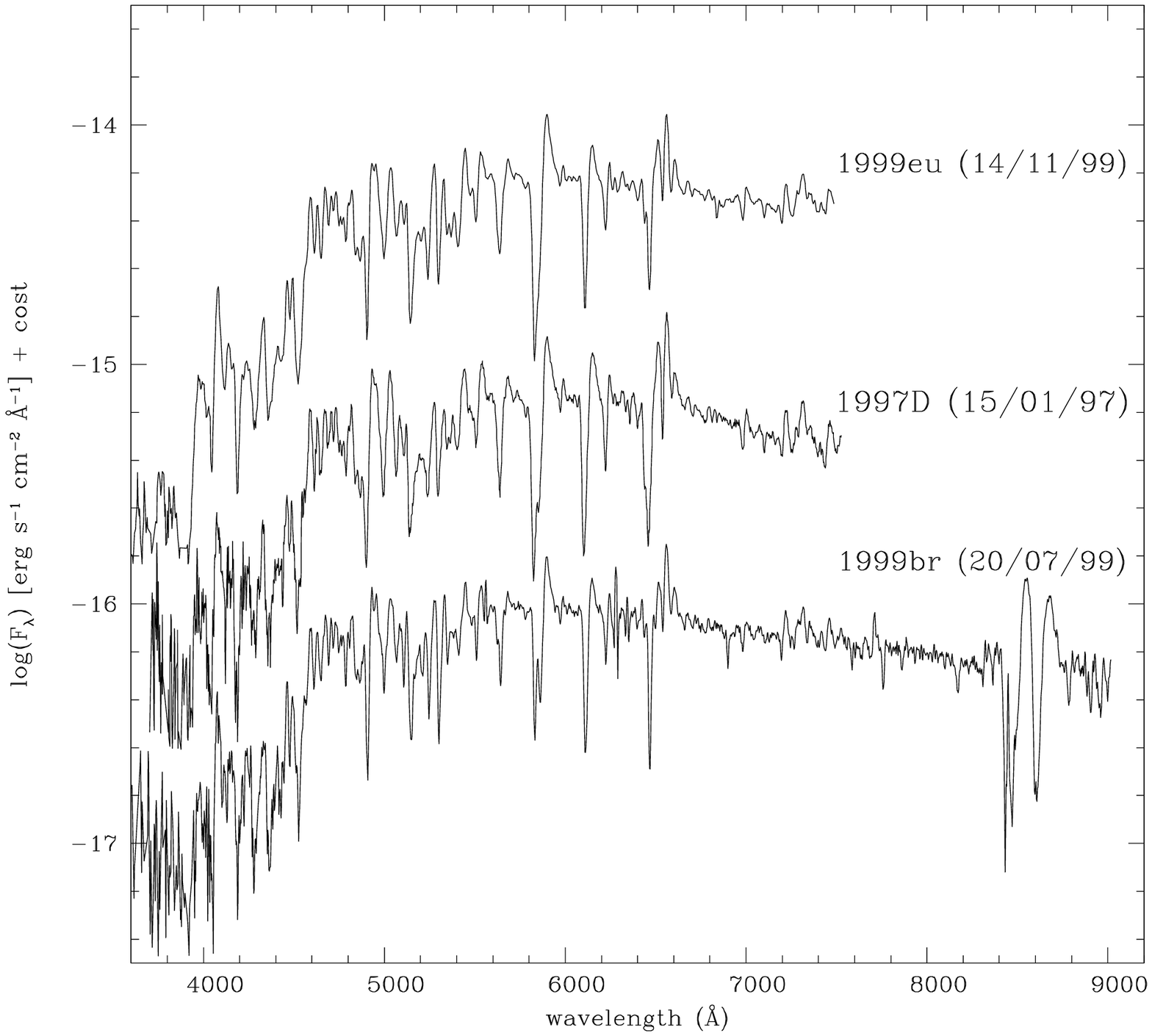}
\caption{Spectra of SN 1999eu, SN 1997D and SN 1999br at about 
the end of the plateau.}
\label{cfr_spec_t0}
\end{figure}

\begin{enumerate}
 \item SN 1999br, the supernova for which we have the best coverage at
 early times, shows the typical features of
 a type~II event (Fig. \ref{spec_ev_99br}).

 The earliest spectrum indicates a phase of about +10~days
 post-explosion, viz. a blue continuum, with relatively broad P--Cygni
 lines of H I and Fe II.  This is consistent with the phase derived
 previously from the light curve. The narrow lines arise from poor
 subtraction of an underlying H II region. During the first month,
 the spectra exhibit P--Cygni lines profiles of H I, Fe II
 (main lines are identified as: multiplets 27, 28, 37, 38  
 between 4100 and 4700 A;
 multiplet 42 at 4924 A, 5018 A and 5169 A; multiplets 48 and 49
 at $\lambda$ between 5200 and 5500 A; multiplets 40, 74, 162, 163 between 
 6100 and 6500 A), Ca II (H$\&$K and the IR triplet), Na I 
 ($\lambda\lambda$ 5890--5896 A), Ti II (many multiplets below $\sim$ 5400 A), 
 Ba II (multiplet 1 at 4554 A, 4934 A; multiplet 2 at 5854 A, 6142 A, 6497 A) 
 and Sc II (e.g. multiplets 23, 24, 28 with the strong line at $\lambda$ 6246 A, 
 29 with a strong blend at $\sim$ 5660 A, 31 with line at $\lambda$ 5527 A). 
 With age the continuum becomes
 redder, and the absorption lines become deeper and less
 blue--shifted.  After a gap of 60~days, the final spectrum was taken
 at about 100~days after explosion, and this shows that a significant
 evolution had taken place in the intervening interval. Numerous
 P--Cygni profiles dominate the spectrum, and their narrow width
 indicates very low expansion velocities ($\sim$ 1000 km s$^{-1}$).
 The most prominent ones are identified as Ca II, Ba II and Sc II (see
 Section 6.3). The spectrum at this stage closely resembles that of SN
 1997D at discovery (see Fig. \ref{cfr_spec_t0}). The line
 identifications at phase $\sim$100~days will be discussed in detail
 in Sect. 6.3.

 \item Four spectra are available for SN 1999eu
 (Fig. \ref{spec_ev_99eu}).  The first two, taken just after
 discovery, are strikingly similar to those of SN 1999br at about
 100~days post--explosion (Fig. \ref{cfr_spec_t0}), i.e. the continuum
 is very red and the lines show expansion velocities of only $\sim$
 1000 km s$^{-1}$. The remaining two spectra, taken during the fast
 post--plateau drop, show even redder continua in agreement with the
 colour evolution discussed above.

 \item The earliest two spectra of SN~1994N (Fig. \ref{spec_ev_94N})
 show the typical appearance of SNe II during the photospheric phase
 \cite{tura94}.  The expansion velocity deduced from the H$\alpha$
 minimum (4500 km s$^{-1}$) is substantially lower than that measured in
 ``normal'' SNe II (see e.g. \ref{vel_lines}), but is similar to those 
 of SN 1999br 3--4 weeks
 post--explosion.  The latest spectrum was taken when the supernova was
 well into its nebular phase, and is reminiscent of the spectrum of SN
 1997D at about 11~months post--explosion \cite{bene01}.  Besides
 H$\alpha$ (FWHM $\approx$ 1200 km s$^{-1}$), the main spectral
 features are the [Ca II] $\lambda\lambda$ 7291--7323~A doublet,
 the Ca II IR triplet, and the strong emission lines of [O~I]
 6300--6364 A.  We also tentatively identify Mg I] 4571 A, H$\beta$,
 Na ID and [Fe II] 7155 A. 
  
 \item Only three spectra of SN 2001dc are available
 (Fig. \ref{spec_ev_01dc}), taken during the plateau phase, at
 $\sim$54, $\sim$91 and $\sim$99 days post--explosion.  The continuum
 is red and there is little evolution in the spectral features during
 this time interval. The last 2 spectra of SN~2001dc resemble those of
 1999br and SN 1999eu at $\sim$ 100 days post--explosion, especially
 the strong and narrow P--Cygni lines of Ba II and Sc II.
 These spectra correspond to
 the beginning of the departure from the plateau phase.  A strong
 P--Cygni line is present in the last spectrum with its emission peak
 at $\sim$ 9245~A.  We tentatively identify this with Sc II
 $\lambda$ 9234~A.

\end{enumerate}

In Fig. \ref{cfr_spec_t0}, we compare the spectra of SN 1997D (1997
January 15), SN 1999br (1999 July 20) and SN 1999eu (1999 November
14).  It can be seen that these are strikingly similar both with
respect to the continua and line components.  We suggest, therefore,
that these spectra were acquired at similar phases (see also Zampieri et al., 2003).  
The light curves of these SNe indicate that this phase was probably about the end of
the plateau phase, with the SN~1999br light curve indicating that this
was $\sim$100~days post--explosion.  For SN~1994N, in spite of the
paucity of observations, comparison of the first spectrum taken on
1994 May 10 with a spectrum of SN 1999br obtained on 1999 May 11,
suggests that SN 1994N was discovered about 1 month after the
explosion.  Indeed the two spectra show the same features and similar
expansion velocities ($\sim$ 2500 km s$^{-1}$, deduced from the
minimum of Fe II 5018 A line).  The adopted epochs of explosion for
all five CC--SNe are given in Tab. \ref{datagal}. These were used in
the analysis of colour and light curves in Sect. 5.

\subsection{The Common Spectro--photometric Evolution of the Low Luminosity SNe II}

\begin{figure*}
\includegraphics[width=19.5cm]{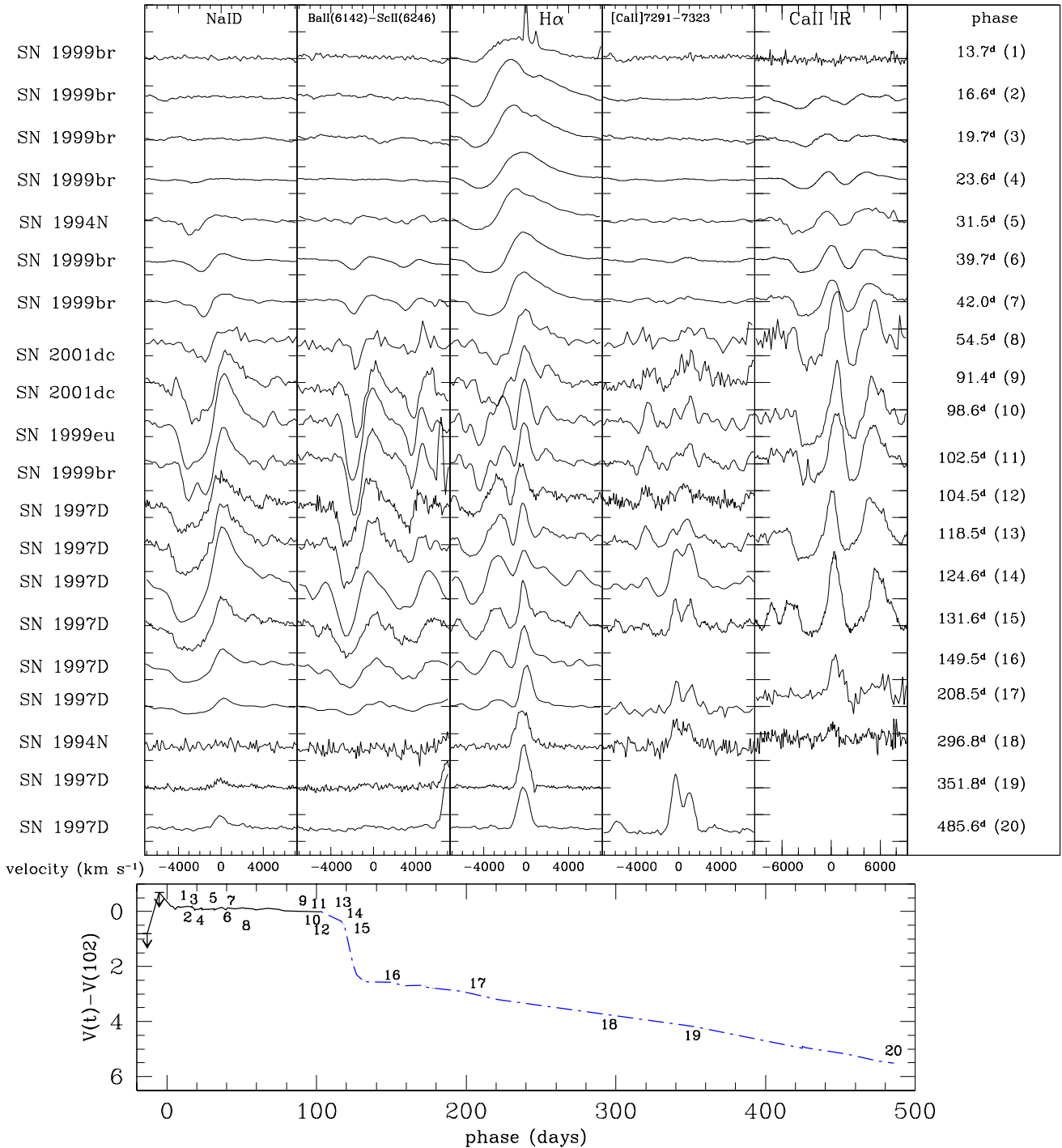}
\caption{Top: evolution of selected regions of the spectra of low
luminosity SNe~II.  The name of the supernova is given in the left
column, while the phase (relative to explosion epoch) is on the
right. The numbers in brackets correspond to the epochs shown on
the schematic light curve (bottom panel), obtained by combining the V
band light curve of SN 1999br (solid curve) and SN 1997D (line--dotted
curve). The magnitudes of both SNe are scaled so that their V
magnitudes match at phase $\sim$ 102 days post--explosion.}
\label{poster}
\end{figure*}

\begin{figure}
\includegraphics[width=8.7cm]{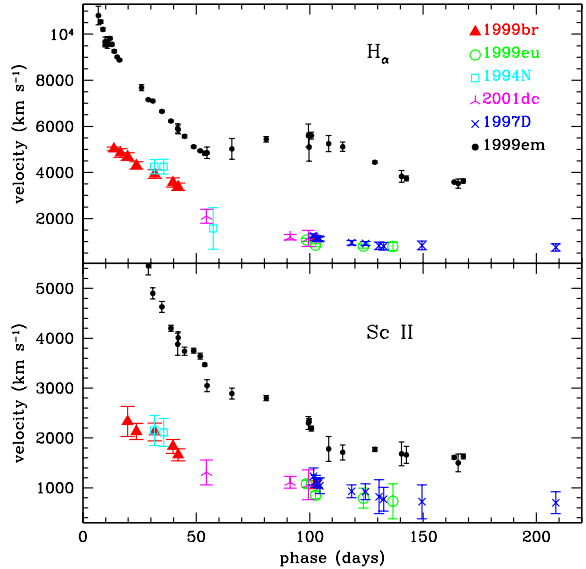}
\caption{Photospheric velocity of low luminosity SNe II deduced from the
minima of P--Cygni absorption of H$\alpha$ and Sc II 6246 A
and comparison with SN~1999em (data from the Padova--Asiago SNe Archive
and Leonard et al., 2002a). The x-axis
shows the phase with respect to explosion.} \label{vel_lines}
\end{figure}
\begin{figure}
\includegraphics[width=8.5cm]{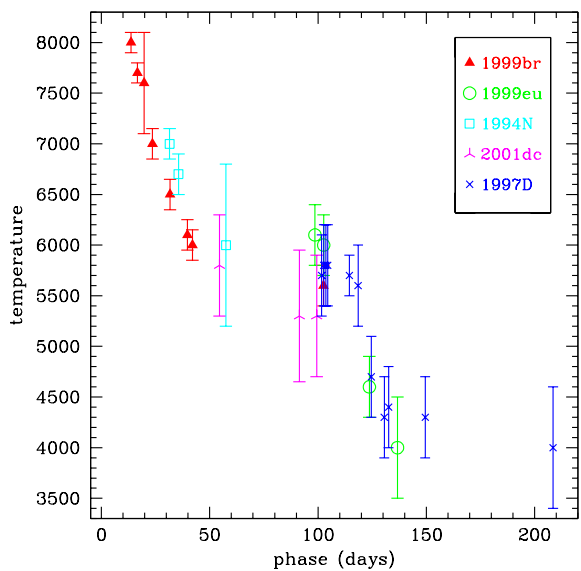}
\caption{Evolution of continuum temperature derived from Planckian
fits to the SNe spectra.  The continuum of SN 2001dc is corrected for
reddening of A$_{B,tot}$ = 1.7, as discussed in Sect. 5. The x-axis
shows the phase with respect to explosion.}
\label{temperature}
\end{figure}

\begin{figure*}
\includegraphics[width=12.085cm]{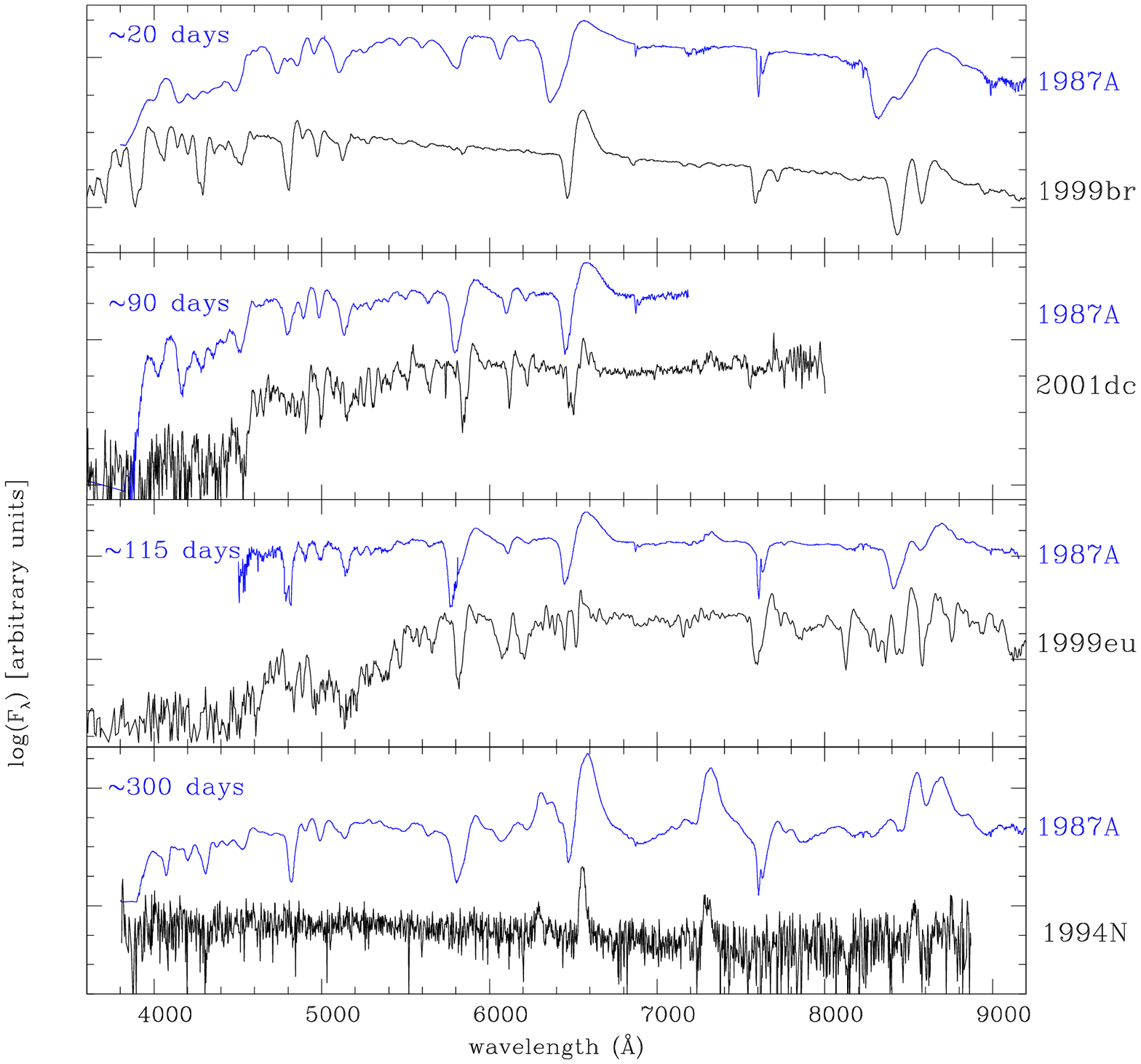}
\caption{Comparison of the spectra of SN 1987A with those of faint SNe
II at similar phases.
The phases with respect to the explosion epoch are shown in the
upper--left corner.} \label{cfr_87A}
\end{figure*}

We have used similarities in the spectra of the low-luminosity SNe to
date the explosions.  In order to better illustrate the properties of
this group of SNe, we now use all the available data to describe the
general spectro--photometric evolution of this class.  In
Fig. \ref{poster} we show the temporal evolution of selected spectral
windows.

The first $\sim$40~days of evolution (spectra 1--8) are
covered mainly by the SN 1999br spectra (cfr. also
Fig. \ref{spec_ev_99br}). The spectra evolve from a continuum
dominated by Balmer lines to a more complex appearance with strong
lines of Na I, Ba II and the Ca II IR triplet. The expansion
velocities are relatively low compared with ``normal'' SNe II, slowing
from 5000 to 3000 km s$^{-1}$ during this early era.  The spectra of
SN~1994N (No. 5) and SN~2001dc (No. 8) fit well into the evolutionary
sequence of SN~1999br.  During this phase, the absorption trough
velocities exhibit rapid movement towards lower velocities (see
Fig. 19 and below).

The next spectrum (No.~9) takes us well into the second half of the
plateau phase.  Clearly the spectroscopic form has changed
significantly in the intervening $\sim$37~days.  The remainder of the
plateau era (spectra 9--12 in Fig. \ref{poster}) is represented by a
single spectrum each from SNe 1997D, 1999br, 1999eu and 2001dc.  As
described earlier, we used the relative similarity of these spectra at
about this epoch to pin down the phases of SNe 1997D and 1999eu.  By
this time (phase $\sim$80--100~days), the absorption troughs of the
more prominent lines indicate relatively low velocities at the
photosphere, e.g. H$\alpha$ and Sc II~6246~A give velocities of
$\sim$1000--1500~km/s (Fig. \ref{vel_lines}). 
In addition, new narrow lines of low excitation
elements like Ba II, Sc II, Fe II, Sr II, Ti II have appeared (see
also Figs. 23--26) and the continuum becomes redder (see e.g. Fig. \ref{spec_ev_99br}).  
In particular the Ba II lines become the strongest features in the spectra.
Note that the contribution of Sr II at short wavelengths is also supported by the detection
of strong lines of multiplet 2 (at 10037 A, 10327 A and 10915 A) in the IR 
spectrum of SN 1997D presented by Clocchiatti et al. (2001).

The evolution up to $\sim$1 month after the plateau (spectra from
about 13 to 16) is characterized by the transition to a nebular
spectral form, with the gradual fading of permitted metallic lines and
P--Cygni profiles. In this phase, narrow forbidden lines have steadily
strengthened (in particular [O I] $\lambda\lambda$ 6300--6364 A and
[Ca II] $\lambda\lambda$ 7291--7323 A, but also some multiplets of [Fe
I], [Fe II] and Mg I]), as reported also in Fig. 5 of Benetti et
al. \shortcite{bene01}.\\ 

The spectra of the latest phase (Nos. 17--20) resemble those of normal
SNe II, with the usual (though narrower) emission features attributed
to forbidden transitions \cite{tura93}.  Nevertheless, the most
prominent line is H$\alpha$, and Na ID is still visible as a faint
emission line after almost 1.5 years.

A comparison between the spectra of the faint SNe~II with those of SN
1987A at comparable phases is shown in Fig. \ref{cfr_87A}.  At the
$\sim$20~day and $\sim$90~day epochs, the main spectral lines are the
same, but the lines of our subluminous CC--SNe have narrower widths
implying lower ejecta expansion velocities.  In addition, at $\sim$90~days the
subluminous SNe show stronger absorption components of Ba II.  As
noted by Turatto et al. (1998) for SN~1997D, this is probably not an
effect of overabundance, but is simply a consequence of a lower ejecta
temperature. 

We conclude that a characteristic property of subluminous CC--SNe is the
very low expansion velocity of the ejecta.
Indeed the photospheric velocities inferred from the absorption minima
of H$\alpha$ and Sc II 6246 A show a common evolutionary path
(Fig. \ref{vel_lines}). During the first $\sim$50~days the line
velocities decrease monotonically from 5000 to 1000--1500 km
s$^{-1}$. Somewhere towards the end of the plateau the velocities
level off at a value $\leq$ 1000 km s$^{-1}$.
At all epochs line velocities are lower than those of the ``typical'' SN II--P
1999em (cf. Fig. \ref{vel_lines}).\\

We have also analyzed the continuum temperature from the spectra of
the SNe by performing a Planckian fit in regions not affected by line
blanketing (e.g. between about 6000 and 8000 A).
The results, shown in Fig. \ref{temperature}, are fairly consistent with a
common evolution.  Starting from a $\sim$ 8000 K at phase $\sim$13
days, the temperature falls reaching a value of 5000--6000~K after
30--40 days.
During the remainder of the plateau the temperature remains almost
constant at $\sim$5500~K.  Presumably this corresponds to the
recombination temperature of the hydrogen as the photophere recedes
through the H--envelope.  The end of the plateau phase is marked by the
onset of a steep temperature decline, settling at T $\sim$ 4000 K as
the supernova enters its nebular phase.

\subsection{Identification of the Spectral Lines}

\begin{figure*}
\includegraphics[width=11.9cm,angle=270]{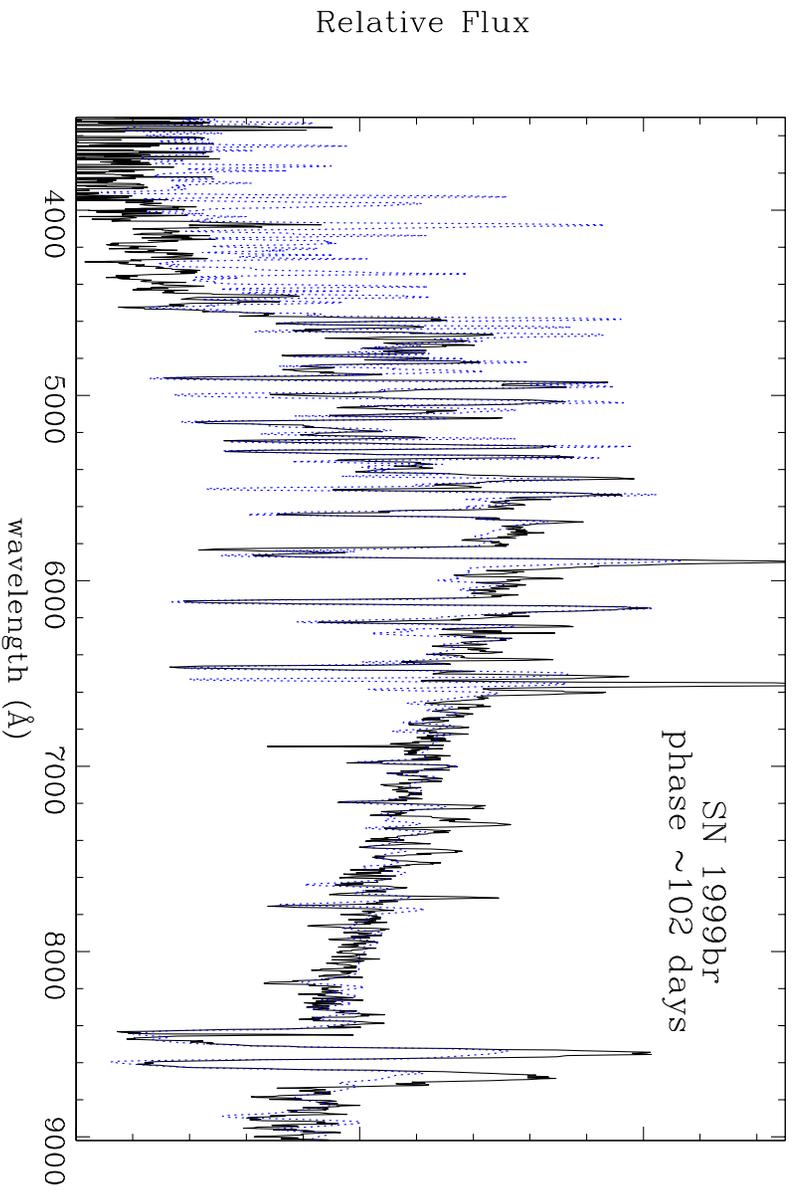}
\caption{Comparison between the observed spectrum of SN 1999br at
phase $\sim$ 102 days and the SYNOW synthetic spectrum. The solid
curve is the observed spectrum, corrected for redshift, the dotted
curve is the synthetic one.} \label{synow1}
\end{figure*}

\begin{figure*}
\includegraphics[width=12.9cm,angle=0]{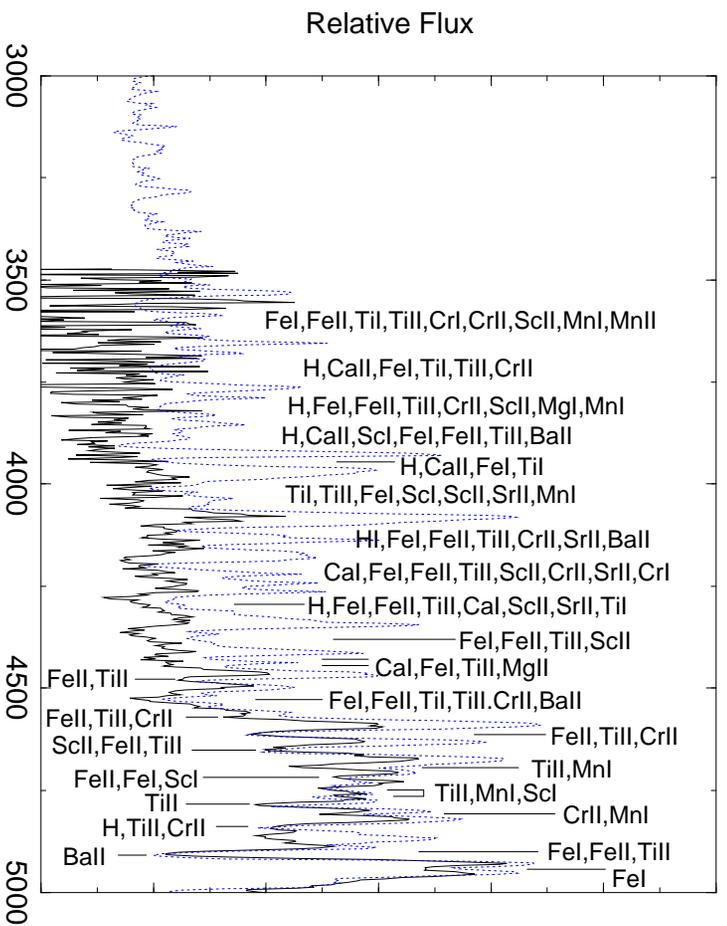}
\caption{Line identification in the wavelength region 3000--5000 A for
the spectra shown in Fig. \ref{synow1}.} \label{synow2}
\end{figure*}

\begin{figure*}
\includegraphics[width=12.9cm,angle=0]{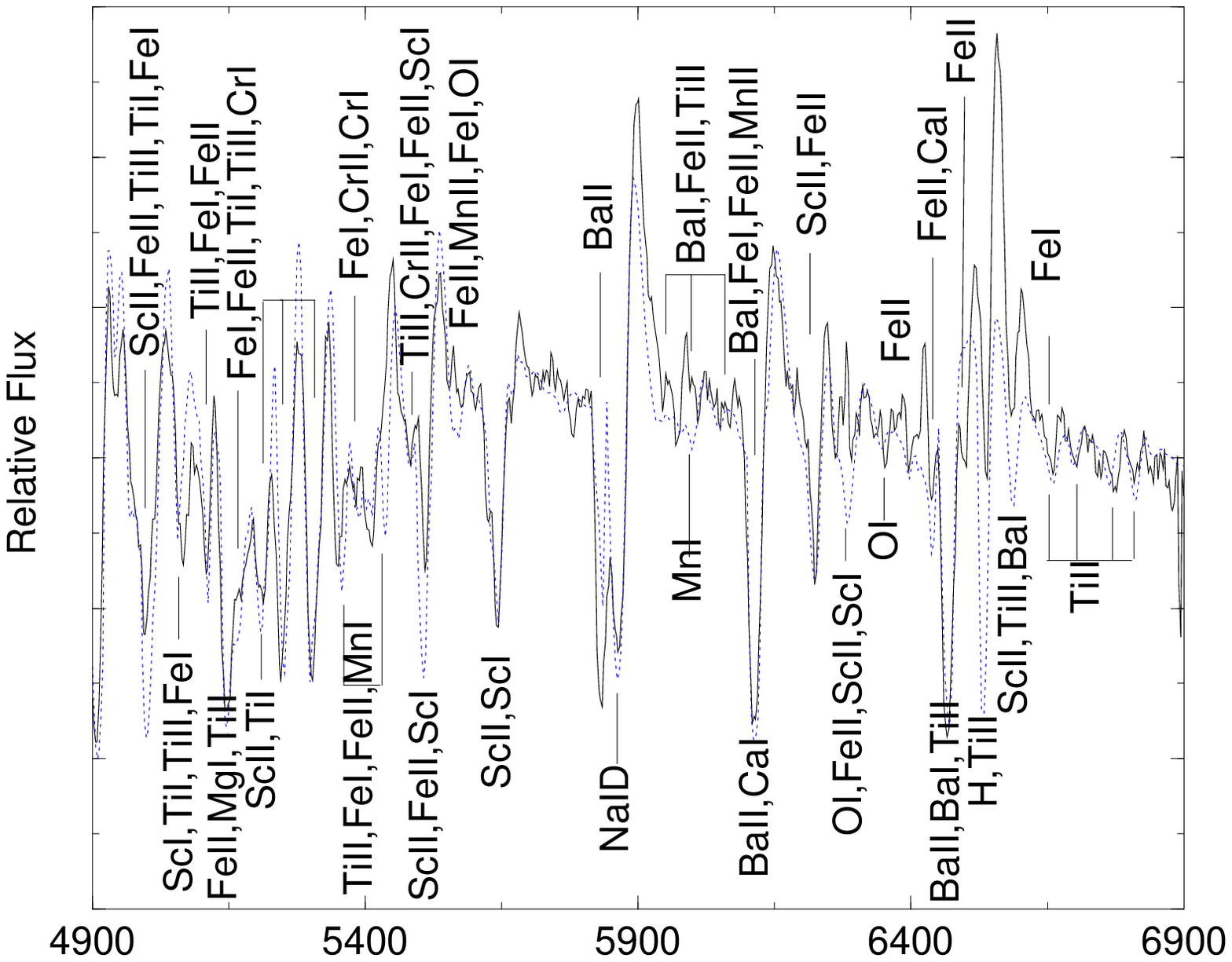}
\caption{Like Fig. 24, but in the wavelength range 4900--6900 A.}
\label{synow3}
\end{figure*}

\begin{figure*}
\includegraphics[width=12.4cm,angle=0]{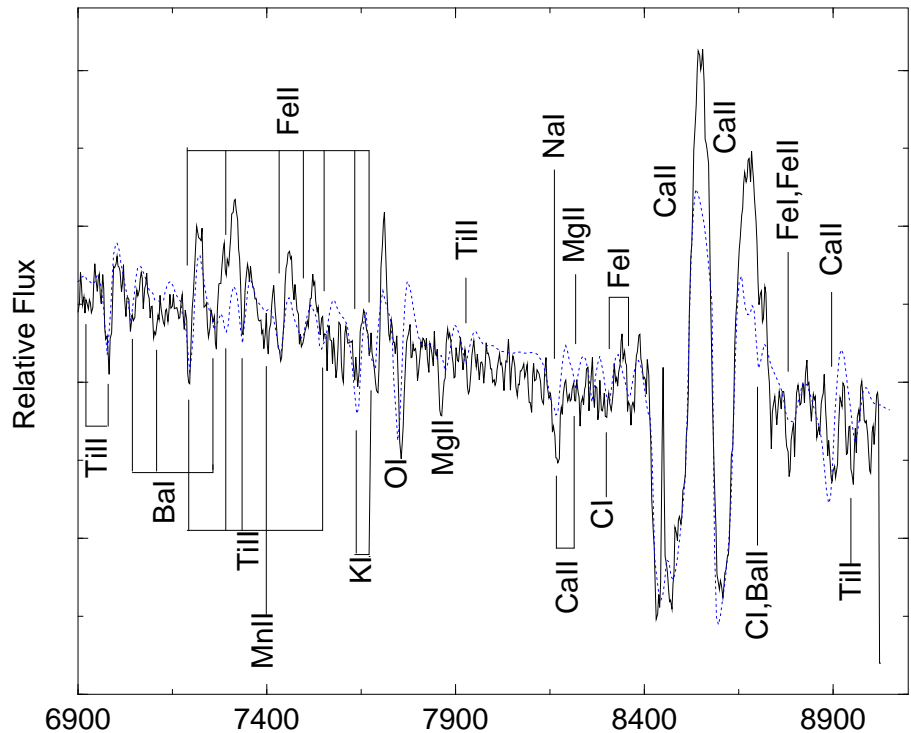}
\caption{Like Fig. 24, but in the wavelength range 6900--9000
A.}\label{synow4}
\end{figure*}

The multiplicity of narrow P--Cygni features in the late photospheric
spectra of the faint SNe II can help us to identify unambiguously the
species responsible. To exploit this opportunity we have modelled the
spectra of SN~1999br at phase 102.5~days.

We used the parameterized code, SYNOW \cite{fish00}, to construct
synthetic photospheric spectra which were then compared with the
observations.  The code works in the Sobolev approximation and
includes a number of simplifying assumptions.  The line--forming
envelope surrounding the continuum--emitting region expands
homologously and with spherical symmetry.  The line source function is
taken to be that of resonance scattering.  The optical depth $\tau$ of
the strongest optical line of each ion is a free parameter. For all
the other lines of a given ion, $\tau$ is found assuming Boltzmann
excitation \cite{jeff90}. 
Other free parameters are the velocity
at the photosphere (v$_{ph}$), the continuum blackbody temperature
(T$_{bb}$), the excitation temperature (T$_{exc}$) of each ion
and the radial
dependency of the optical depths.  We adjusted these parameters to
produce a plausible spectral match to the $\sim$102~day spectrum of SN
1999br.  In the final model we adopted a power--law radial dependency
of index $n$ = 4 with T$_{bb}$ = 5600 K and v$_{ph}$ = 970 km
s$^{-1}$; T$_{exc}$ varies among the ions, but it is about 4000--5000
K for neutral ions and 8000--10000 K for ionized species.\\

The resulting spectral model, obtained including 22 different ions, 
is compared with the whole observed
spectrum in Fig. \ref{synow1}, and with expanded sections shown in
Figs. \ref{synow2}--\ref{synow4}.
The main spectral features are generally quite well reproduced. In
addition the strong line--blanketing behaviour in the blue ($\lambda
\leq$ 4600~A) is, at least qualitatively, also seen in the model
(Fig. \ref{synow2}).\\  
The line--blanketing is due to the presence of many strong metallic
lines, including contributions from neutral and singly--ionized ions of
Fe, Ti, Cr, Sc, Mn, Ca and Sr.  The Sr II lines are particularly deep,
while the Balmer lines are fainter than is observed in normal SNe
II.\\
The region between 4600 and 6500~A (Fig. \ref{synow3}) is dominated by
lines of Fe II, Ti II, Sc II and Ba II, although contributions from
lines of neutral ions cannot be excluded.  At this phase the
absorption lines of Ba II are exceptionally strong, exceeding in depth
the Na ID and H$\alpha$ lines.  The reproduction of the spectrum in
the H$\alpha$ region is poor due to the inadequacy of the pure
resonance--scattering assumption for the H$\alpha$ line.
Nevertheless, to reproduce the complex profile in this region, we
require the simultaneous presence of Ba II, H$\alpha$ and Sc II.\\
On the red tail of H$\alpha$ up to $\sim$7700~A (Fig. \ref{synow3}
and Fig. \ref{synow4}) many narrow, faint Ti II and Fe II lines are
superimposed on a strong continuum. However, also present may be
growing contributions from nebular emission lines, in particular [Fe
II] and [Ca II] $\lambda\lambda$ 7291, 7324 A.  As we move still further
to the red, the spectrum becomes dominated by the Ca II IR
features. Other tentatively identified lines include: K I, O I, Mg II,
Na I, C I, Fe I, Fe II, Ba II and Ti II (Fig. \ref{synow4}).\\
 
\section{Discussion}

Following our determination of the explosion epochs, we can conclude
that our observations are consistent with low--luminosity SNe having
plateaux of duration $\sim$100~days, similar to that seen in more
normal SN~II--P events. This may constitute evidence for the presence of
massive envelopes around low--luminosity SNe \cite{zamp03}.
We attribute a $\sim$100~day plateau to SN~1997D.  This is
significantly longer than that estimated by Turatto et
al. \shortcite{tura98} and Benetti et al.  \shortcite{bene01}.
In general, the plateau luminosities of the faint CC--SNe are unusually low, and show a range
of values between the different events spanning more than 1~magnitude
(see Sect. 5, Fig. \ref{abs_V} and Fig. \ref{bolom}).  Although
spectroscopic coverage is incomplete for individual events, the data
are consistent with the low--luminosity SNe having a similar
spectroscopic evolution, characterised by unusually narrow line
widths. \\

At the end of the plateau the light curves show a steep decline,
typically 3 magnitudes in 30--50 days. This effect has also been
observed in other type II--P SNe [e.g. SN 1994W, Sollerman et al. (1998);
SN 1999em, Elmhamdi et al. (2003)].  Thereafter, for 3 of the 4 SNe for
which we have post--plateau light curves, the decline rate is consistent
with the power source being the radioactive decay of $^{56}$Co. The
exception is SN~1999eu, discussed below.
Assuming complete trapping of the $\gamma$--rays, we compare the 
{\sl ``OIR''} pseudo--bolometric luminosities (Fig. \ref{bolom}) 
with those of SN~1987A at similar epochs and, using the relation
M$_{SN}$(Ni) = 0.075 $\times$ $\frac{L_{SN}}{L_{87A}}$ M$_{\odot}$, we obtain the
following useful lower limits for the ejected $^{56}$Ni masses: $\sim$
0.008M$_\odot$ for SN~1997D, $\sim$0.002 M$_{\odot}$ for
SN~1999br, $\sim$0.006 M$_{\odot}$ for SN~1994N (using only the {\sl
BVR} bands), $\sim$0.006 M$_{\odot}$ for SN 2001dc (with a large
uncertainty due to the poor late time photometry).  We note that the
value for SN~1997D is a factor 4 higher than the value estimated by
Turatto et al., 1998, due to differences in estimated explosion epoch
and distance (see Tab. \ref{datagal} for new estimates). 
By analogy with SN~1987A, we can deduce that
no more than 50$\%$ of the luminosity was lost in unobserved
bands.  Consequently we may conclude that a common feature of these
events is ejection of a very low $^{56}$Ni mass.\\

As mentioned above, SN~1999eu is noticeably different from either
SN~1997D or SN 2001dc in that it exhibits a particularly large
post--plateau decline of about 5~magnitudes.
The large decline refers to the entire duration of the observations prior
to the disappearance of SN~1999eu behind the sun.
It's remarkable that a similar behaviour was already observed in SN~1994W
\cite{soll98}. 
Moreover, its luminosity
drops to a level which is about 3 times lower than would be expected
from a simple backward extrapolation of the later $^{56}$Co
radioactively--driven tail. 
Two possible explanations present
themselves.  One is the early condensation of dust in the ejecta,
causing much of the luminosity to be shifted into the unobserved mid--
and far--infrared. Then as the ejecta expanded, the dust became
optically transparent to the optical luminosity and so the light curve
increasingly followed the radioactive decline rate, with a luminosity
corresponding to 0.003 M$_{\odot}$ of $^{56}$Ni.  A similar phenomenon
was seen in SN~1987A but at a later epoch.  Difficulties with the dust
hypothesis are: (a) the rather early epoch of the condensation, while
the temperature is still rather high in at least some of the ejecta;
(b) its rather ``convenient'' coincidence with the decline at the
end of the plateau.  Unfortunately, nebular spectra of SN 1999eu are
not available and so we cannot check for the occurrence of a red--wing
truncation that dust condensation might be expected to produce
(cf. SN~1987A and SN~1999em).  The second possibility is that the
late--time tail is, in fact, powered mostly by ejecta--CSM interaction
and not by radioactive decay. This would imply an extremely low
ejected mass of $^{56}$Ni ($<$0.001 M$_{\odot}$).  A possible
difficulty is that if the light curve is being mostly driven by such
an interaction, then there is no obvious reason why it should be close to
the decay rate of $^{56}$Co between 350 and 550 days.\\

In summary, the group of five CC--SNe considered here are similar to
more typical CC--SNe in that a clear plateau phase occurs lasting for
$\sim$100~days, followed by a late--time decline driven by the decay of
$^{56}$Co.  A similar temperature behaviour is seen
(Fig. \ref{temperature}), and the identities of the spectral lines at
all phases are also typical.  However, these SNe differ in that (a)
during the plateau phase the luminosity is at least a factor
10 times less than found in typical CC--SNe, (b) the expansion
velocity is unusually slow (Fig. \ref{vel_lines}), both during the
photospheric and nebular phases, and (c) the mass of $^{56}$Co which
drives the late--time tail is at least a factor $\sim$10 lower than
normal. \\

A key issue to address is whether or not the five low--luminosity
CC--SNe are members of a separate class of CC--SNe arising from a
distinctly different progenitor type and/or explosion process, or are
they simply samples from the extreme low--luminosity tail of a
continuous distribution of otherwise normal explosions.
We are currently investigating this problem through
the study of SNe~II having luminosities intermediate between normal
objects (like SN 1988A or SN 1969L) and the faint events discussed
here.  Hamuy \shortcite{ham03a} has already considered 24
type~II plateau SNe, deriving a wide range of plateau luminosities
($\sim$5 magnitudes), expansion velocities ($\sim\times$5) and ejected
$^{56}$Ni masses ($>\times$150). His sample includes two of our
low--luminosity events viz. SNe 1999br and 1997D. He found that both
the mid--plateau M$_V$ and ejecta velocity correlate with the mass of
$^{56}$Ni ejected as deduced from the exponential tail. For a subset
of 16 SNe II--P the explosion energy and total ejected mass
also correlate with the observed properties of plateau luminosity and
velocity.  Hamuy asserts that the physical properties of SNe II--P
exhibit a continuous range of values, and concludes that SNe II--P
form a one--parameter family.  We note that the four low--luminosity
SNe for which we have a fair estimate of M$_V$ follow the trend of
M$_V$ at 50~days versus M(Ni$^{56}$) shown in Hamuy's Fig.~3. It
therefore seems reasonable to incline towards the single continuous
distribution scenario.  Given Hamuy's correlation between ejecta
kinetic energy and mass of $^{56}$Ni ejected, this implies that
low--luminosity CC--SNe are also low--energy events.  
Similar results for a large sample of SNe II--P support Hamuy's conclusion 
that a continuous trend of physical parameters in SNe II--P exists, 
also including extremely low and moderately
low luminosity events  (Zampieri et al., in prep.).  \\

To account for the low--luminosity CC--SNe, a natural approach is to
examine the extreme ends of the mass spectrum of progenitors from
which SNe II--P are believed to arise.  The high--mass scenario has been
proposed by Turatto et al. (1998), Benetti et al. (2001) and
Zampieri et al.  (2003). Turatto et al. (1998) found that the observed
behaviour of SN~1997D could be reproduced by the explosion of a
progenitor with $M > 25 M_\odot$, radius $R \leq 300 R_\odot$ and
explosion energy $E\approx 4\times 10^{50}$ erg. They also suggested
that the central remnant was a black hole (BH) and that significant
fallback of stellar material onto the collapsed remnant may have taken
place.  Zampieri et al. \shortcite{zamp03} studied the behaviour of
both SN~1997D and SN~1999br.  They
found that the light curves and the evolution of the continuum
temperature and expansion velocity are well reproduced by a
comprehensive semi--analytical model in which the envelope is $\le$
10$^{13}$ cm (140 R$_{\odot}$), the ejecta masses are $\sim$ 14--20
M$_{\odot}$ and explosion energy is $E\leq 10^{51}$ erg. They deduced
somewhat lower progenitor masses than did by Turatto et al., estimating 
a progenitor with mass $\geq$ 19 M$_\odot$ for SN 1997D 
and $\geq$ 16 M$_\odot$ for SN 1999br.
From the large ejected masses and low explosion energy, Zampieri et al. 
conclude that SN 1997D and SN 1999br are intermediate mass, BH--forming CC--SNe.\\

A contrary conclusion was reached by Chugai $\&$ Utrobin (2000).  In
their analysis of the early-- and late--time spectra of SN~1997D, they
found that model spectra for a supernova resulting from a 24~M$_\odot$
progenitor were incompatible with the observed nebular spectra.  They
considered also a low--mass progenitor model in which the observed
behaviour of SN 1997D is reproduced by the low energy explosion
($E\approx 10^{50}$ erg) of a low--metallicity progenitor. The star
has a radius R $\leq$ 85 R$_{\odot}$ and a total main sequence mass of
M = 8--12 M$_{\odot}$ (of which 6 M$_{\odot}$ becomes H rich ejecta).
In this case, the remnant is expected to be a neutron star. They found
that this model gave more successful reproduction of the spectra at
both early and late times.
However, in their analysis Chugai $\&$ Utrobin assumed a plateau duration of 
$\sim$60 days. The new observational evidence provided here and
in Zampieri et al. (2003) indicates that the plateau of low--luminosity SNe 
is $\ga$ 100 days. Therefore, it is not clear if the low mass model may 
still be able to reproduce the observations with such a long plateau duration.\\

Incidentally, Cappellaro (in prep.) find that roughly 50$\%$ of
low--luminosity SNe II--P occur in late type (Sc) galaxies where recent
star formation has occurred, whereas the fraction for normal SNe II is
only $\sim$ 15 $\%$. This favours a high--mass progenitor for the
low--luminosity events. 

Ultimately, it may be possible to devise new observational tests for
the high--mass scenarios.  Woosley \& Weaver (1995) have
shown, for a low energy explosion occurring in a massive progenitor, a
considerable fraction of $^{56}$Ni and other heavy elements may indeed
remain gravitationally bound to the compact core and fall back onto
it.  If the fallback is too large it can lead to problems in
reproducing the observed Barium and other $r$-process elements (Qian
\& Wasserburg, 2001). However, hydrodynamic calculations show that 1
$M_\odot$ of stellar material can easily remain bound and accrete onto
the core, forming a BH. 
In principle one might detect such a BH via the late--time light curve when accretion
luminosity induced by the fallback overcomes the radioactive decay
luminosity (Zampieri, Shapiro \& Colpi, 1998).  For spherical symmetry,
the accretion luminosity is predicted to decay as $t^{-25/18}$
\cite{zam98b}, but it is also not expected to dominate the radioactive
luminosity until $\sim$ 3 years post--explosion (Balberg et al., 2000).
Consequently this effect is currently at the sensitivity limit of the
largest ground--based telescopes.

\section{Are low--luminosity SNe rare?}

A significant fraction of all types of SNe may be underluminous, but
their faintness may produce a statistical bias against the discovery
of such events \cite{scha96}.  Richardson et al.  \shortcite{rich01}
found that for SNe in our
Galaxy and in nearby galaxies ($\mu \leq$ 30), possibly more than
20$\%$ have an intrinsic M$_B$ fainter than --15.  The majority of
these events are of type II--P (both 1997D--like and 1987A--like
events).  Chugai $\&$ Utrobin (2000) speculate that the Galactic
supernovae of 1054 and 1181 might have been sub--luminous events
similar to SN 1997D.  They estimate that such faint SNe make up $\sim$
20$\%$ of all SNe II--P.

The analysis of archival data and published material suggests that
several other 1997D--like events have been observed in the past.  For
example, SN 1973R (Ciatti $\&$ Rosino, 1977) exhibited low plateau
luminosity, relatively narrow P--Cygni lines and a light and colour
curve behaviour similar to that of SN~1997D, suggesting that it
belongs to this group of sub--luminous SNe.  Other candidates might be
SN 1923A \cite{pata94}, SN 1978A \cite{elli78}, SN 1999gn (VSNET data;
Dimai $\&$ Li, 1999; Ayani $\&$ Yamaoka, 1999), SN 2000em
\cite{stro00}, SN 2001R (Weisz et al., 2001; Matheson et al., 2001)
and the recent SN 2003Z (Qiu $\&$ Hu, 2003; Matheson et al. 2003).  We
note, in particular, that the spectrum of SN 2000em\footnotemark[4] 
taken by the NGSS Team \footnotetext[4]{http://www.ctio.noao.edu/$^{\sim}$ngss/ngss4/ngss4.html}
shows many features that are reminiscent of those observed in
SN~1997D. The H$\alpha$ emission dominates with respect to the Ba II
6497 A line, thus resembling e.g. spectra (8) and (9) in
Fig. \ref{poster}.

SN 1978A \cite{elli78} exhibited narrow P--Cygni spectral features
resembling those of SN 1997D. A complex sequence of narrow absorption
and emission lines are visible in a rather noisy spectrum 
(resolution $\sim$ 10 A).
SN 1978A is also peculiar in that while its late--time luminosity was
similar to that of 1997D--like events, at discovery its absolute
magnitude was much higher: $\sim$ --18.6 [Gilmore \shortcite{gilm78}; 
Zealey $\&$ Tritton \shortcite{zeal78}].  
A possible explanation could be an early
interaction with a thin circumstellar shell ejected just before the SN
explosion, powering the luminosity at very early phases. However, in
this scenario it is difficult to explain the observed low expansion
velocities at just 2--3 weeks post--explosion.  SN 1994W \cite{soll98}
is also a case where the luminosity at maximum was high, but where a
low (M$_{Ni} \leq$ 0.015 M$_{\odot}$) or very low (M$_{Ni} \leq$ 0.0026 M$_{\odot}$) 
mass of radioactive \ni\/ was
ejected, depending on the contribution to the luminosity of an
ejecta--CSM interaction. \\

The discovery in the past few years of a few SNe spectroscopically
classified as 1997D--like events, suggests that the rate of such events
is significant.  Among type II SNe discovered during 1992--2001 in a
volume--limited sample (recession velocities $\leq$ 4000 km s$^{-1}$)
[Barbon et al., 1999, and recent updates], at least 5 are
spectroscopically similar to SN 1997D.  This suggests that the incidence
of such very low--luminosity events is about 4--5$\%$ of all type II SNe.

\section{CONCLUSIONS}

In this paper we have presented photometric and spectroscopic
observations of 4 new SNe II, viz. SNe 1999br, 1999eu, 1994N and
2001dc.  Together with SN 1997D, we have shown that they form a group
of exceptionally low--luminosity events.  The discovery of these faint
SNe is difficult and their monitoring requires in general large
telescopes and long integration times.  Partly in consequence of this,
the temporal coverage of individual supernovae was erratic and
incomplete.  However, taken together, they suggest a fairly
homogeneous set of properties, and provide a reasonably complete
picture of the photometric and spectroscopic evolution of this type of
supernova.  Establishment of the phase of these events was important,
and the best--observed SN~1999br was particularly valuable in this
regard (see also Zampieri et al., 2003).

We find that the group of five CC--SNe considered here are similar to
more typical CC--SNe in that a clear plateau phase occurs lasting for
$\sim$100~days, followed by a late--time decline driven by the decay of
$^{56}$Co.  A similar temperature behaviour is also seen. 
The spectrum evolves from a relatively
normal SN~II photospheric spectrum to one characterized by narrow
lines (v $\approx$ 1000 km s$^{-1}$), a red continuum and strong Ba II
lines.  These SNe are unusual in that during the plateau phase the
luminosity (both V--band and bolometric) is at least a factor
10 times less than found in typical CC--SNe. In addition, the
expansion velocity is unusually slow both during the photospheric and
nebular phases.  At the earliest epochs, the photosphere is located at
layers expanding at $\sim$ 5000 km s$^{-1}$, but within two months it
recedes to v $\sim$ 1000 km s$^{-1}$.  The mass of $^{56}$Co which
drives the late-time tail is at least a factor $\sim$10 lower than
normal, indicating ejected masses of $^{56}$Ni in the range 2--8
$\times$ 10$^{-3}$ M$_{\odot}$.

Comparison with a sequence of more normal SNe~II (Hamuy, 2003; 
Zampieri et al., in preparation) suggests that this group of SNe represent 
the extreme low--luminosity tail of a single continuous distribution. This 
indicates that the low--luminosity CC--SNe are also low--energy events.  
Although evidence for a low--mass progenitor of low--luminosity
CC--SNe may not be completely ruled out yet, recent work seems to
support a high mass progenitor scenario (Zampieri et al., 2003).

Finally, we note that selection effects probably limit the rate of 
discovery of low--luminosity SNe, and that their true incidence may 
be as high as 4--5$\%$ of all type II SNe.

\section*{Acknowledgments}
We acknowledge support from the MIUR grant Cofin 2001021149,
and partial support from NSG grants AST--9986965 and AST--0204771, and
NASA grant NAG5--12127.\\ 
We are grateful to M. Hamuy \& M. M. Phillips for providing their data
for SN 1999br before publication and to the referee D. C. Leonard for helpful
comments which have improved the paper.
We thank D. Bramich, L. Contri, R. Corradi, S. Desidera, P. Erwin,
B. Garcia, E. Giro, D. Lennon, L. Lessio, N. O'Mahoney, J. Rey and
A. Zurita for technical support and assistance in the observations of
SN~2001dc.
A. P. thanks the University of Oklahoma for generous hospitality.\\
We are grateful to the VSNET observers, in particular for the useful
information about SN 1999eu.\\
This research has made use of the LEDA database
and of the NASA/IPAC Extragalactic
Database (NED) which is operated by the Jet Propulsion Laboratory,
California Institute of Technology, under contract with the National
Aeronautics and Space Administration.

\end{document}